\newif\iffigs\figsfalse

\iffigs \input epsf \else  \fi

\documentstyle[preprint,aps,floats,eqsecnum]{revtex}
\begin{document}
\newlength{\pubnumber} \settowidth{\pubnumber}{IASSNS-HEP-93/999}

\relax
\citation{email}
\citation{T1}
\citation{ALT}
\citation{T1}
\citation{KT}
\citation{ZFpf,ZFft}
\citation{NST}
\newlabel{sI}{{\uppercase {i}}{2}}
\citation{T1}
\citation{ZFft}
\newlabel{sII}{{\uppercase {ii}}{3}}
\newlabel{fsca}{{2.1}{3}}
\newlabel{lamb}{{2.2}{3}}
\newlabel{autom}{{2.3}{4}}
\newlabel{untwistbypass}{{2.4}{4}}
\newlabel{fig1}{{1}{4}}
\newlabel{dblbypass}{{2.5}{4}}
\citation{ZFft}
\citation{ZFft}
\newlabel{twistedom}{{2.6}{5}}
\newlabel{twcurbp}{{2.7}{5}}
\newlabel{ggtequiv}{{2.8}{5}}
\citation{ZFpf,ZFft}
\newlabel{Rbypass}{{2.9}{6}}
\newlabel{ZFmodiden}{{2.10}{6}}
\newlabel{twistint}{{2.11}{6}}
\newlabel{Ddefn}{{2.14}{6}}
\newlabel{fig2}{{2}{7}}
\newlabel{RsectGCR}{{2.16}{7}}
\newlabel{GLcom}{{2.17}{7}}
\citation{ZFpf}
\citation{T1}
\citation{T1}
\newlabel{Rsectzmalg}{{2.18}{8}}
\newlabel{GGtope}{{2.19}{8}}
\citation{T1}
\newlabel{ohtwo}{{2.20}{9}}
\newlabel{twistGCR}{{2.21}{9}}
\newlabel{twistGCR3}{{2.22}{9}}
\newlabel{twistGCR2}{{2.23}{9}}
\newlabel{sIII}{{\uppercase {iii}}{9}}
\citation{NST}
\newlabel{Asum}{{3.1}{10}}
\newlabel{pscR}{{3.2}{10}}
\newlabel{ApictR}{{3.3}{10}}
\citation{T1}
\newlabel{ApictRan}{{3.5}{11}}
\newlabel{twiscom}{{3.6}{11}}
\newlabel{spcase}{{3.7}{11}}
\newlabel{Apict2R}{{3.8}{11}}
\newlabel{Apict3R}{{3.9}{11}}
\newlabel{spur}{{3.10}{12}}
\newlabel{spstde}{{3.12}{12}}
\newlabel{ApictRprime}{{3.14}{13}}
\citation{T1}
\newlabel{pscRprime}{{3.17}{14}}
\newlabel{sIV}{{\uppercase {iv}}{14}}
\citation{T1}
\citation{orbs}
\citation{Itoh}
\newlabel{covflds}{{4.2}{15}}
\newlabel{Grepd}{{4.3}{15}}
\newlabel{Grepspltil}{{4.4}{15}}
\newlabel{Btwistby}{{4.5}{15}}
\newlabel{cfiveops}{{4.6}{15}}
\citation{Itoh}
\citation{Itoh}
\newlabel{Btwmodeexp}{{4.7}{16}}
\newlabel{betacomm}{{4.8}{16}}
\newlabel{narain}{{4.10}{16}}
\newlabel{twvtx}{{4.11}{16}}
\newlabel{Bbope}{{4.12}{17}}
\newlabel{sbpr}{{4.13}{17}}
\newlabel{twvtxmodexp}{{4.14}{17}}
\newlabel{so21modiden}{{4.16}{17}}
\newlabel{gammazero}{{4.17}{18}}
\newlabel{gammalg}{{4.18}{18}}
\newlabel{dimcomp}{{4.19}{18}}
\newlabel{Xbosexp}{{4.21}{18}}
\newlabel{Gtwmodexp}{{4.22}{19}}
\newlabel{epgcr}{{4.24}{19}}
\newlabel{gtraceiden}{{4.25}{19}}
\newlabel{stoep}{{4.27}{19}}
\newlabel{ttoep}{{4.28}{20}}
\newlabel{Ltwmodexp}{{4.29}{20}}
\newlabel{pscT}{{4.30}{20}}
\newlabel{tauzero}{{4.32}{20}}
\newlabel{interel}{{4.33}{20}}
\newlabel{constraint}{{4.35}{21}}
\newlabel{polconstr}{{4.37}{21}}
\newlabel{tauone}{{4.38}{21}}
\newlabel{interel2}{{4.39}{21}}
\citation{T1}
\newlabel{sV}{{\uppercase {v}}{22}}
\citation{ALyT}
\newlabel{RintrelGCR}{{5.1}{23}}
\newlabel{diracompat}{{5.2}{23}}
\newlabel{c2Grep}{{5.4}{23}}
\newlabel{Rtauzero}{{5.10}{24}}
\newlabel{Rinterel}{{5.11}{25}}
\newlabel{Rtauone}{{5.12}{25}}
\newlabel{Rinterel2}{{5.13}{25}}
\citation{orbs,asymm}
\newlabel{Bdegen}{{A2}{26}}
\newlabel{Bzmreps}{{A4}{26}}
\newlabel{Bsoalg}{{A5}{26}}
\newlabel{Btwzmod}{{A6}{26}}
\newlabel{zmreps}{{A11}{27}}
\bibcite{email}{*}
\bibcite{T1}{1}
\bibcite{ALT}{2}
\bibcite{KT}{3}
\bibcite{ZFpf}{4}
\bibcite{ZFft}{5}
\bibcite{NST}{6}
\bibcite{orbs}{7}
\bibcite{Itoh}{8}
\bibcite{ALyT}{9}
\bibcite{asymm}{10}

\preprint{\parbox{\pubnumber} {\begin{flushleft}
IASSNS-HEP-93/58\\ CLNS-93/1251\\
hep-th/9311131\\ \end{flushleft}}}
\draft
\tighten
\title{ Tree scattering amplitudes of the\\
spin-4/3 fractional superstring II:\\
The twisted sectors}
\author{Philip C. Argyres\cite{email}}
\address{School of Natural Sciences, Institute for Advanced Study,
Princeton, NJ 08540}
\author{S.-H. Henry Tye}
\address{Newman Laboratory of Nuclear Studies, Cornell University,
Ithaca, NY 14853}
\maketitle
\begin{abstract}
The spin-4/3 fractional superstring is characterized by a world-sheet
chiral algebra involving spin-4/3 currents.  The discussion of the
tree-level scattering amplitudes of this theory presented in
Ref.\ \cite{T1} is expanded to include amplitudes containing two
twisted-sector states.  These amplitudes are shown to satisfy spurious
state decoupling.  The restriction to only two external twisted-sector
states is due to the absence of an appropriate dimension-one vertex
describing the emission of a single twisted-sector state.  This is
analogous to the ``old covariant'' formalism of ordinary superstring
amplitudes in which an appropriate dimension-one vertex for the
emission of a Ramond-sector state is lacking.  Examples of tree
scattering amplitudes are calculated in a $c=5$ model of the spin-4/3
chiral algebra realized in terms of free bosons on the string
world-sheet.  The target space of this model is three-dimensional flat
Minkowski space-time and the twisted-sector physical states are
fermions in space-time.  Since the critical central charge of the
spin-4/3 fractional superstring theory is 10, this $c=5$ model is not
consistent at the string loop level.
\end{abstract}
\narrowtext\noindent

\section{Introduction and summary}
\label{sI}

Fractional superstrings \cite{ALT} are string theories whose physical
state conditions are generated by fractional-spin currents on the
world-sheet.  The resulting
critical central charges of these strings are found
to be less than that of the superstring, suggesting
the possibility of string theories with critical space-time
dimensions less than 10.  This paper is the second of two papers
examining the properties of the simplest such fractional
superstring, namely the one with spin-4/3 currents on the
world-sheet.  The first paper in this series \cite{T1} described
tree scattering amplitudes of this string
for states in the untwisted sectors of the spin-4/3 fractional
superconformal algebra, and showed that they obey duality and
spurious state decoupling.  Also, a specific $c=5$ (non-critical)
model of this string with a three-dimensional space-time
interpretation was constructed, where it was shown
that the untwisted-sector physical states correspond to
space-time bosons including the graviton and Yang-Mills
bosons.  In Ref.\ \cite{KT} a no-ghost theorem for these
$c=5$ untwisted-sector states was proven.

This paper examines the twisted sectors
of the spin-4/3 fractional superstring.  In Section \ref{sII} we define
the twisted sectors in terms of the monodromies of
the fractional currents with states in those
sectors.  We derive general properties that models
having twisted sectors must obey using the methods of
Refs.\ \cite{ZFpf,ZFft}.  It is found that there are
two types of twisted sectors, which we denote by
$R$ and $R'$, whose occurence depends on how a ${\bf Z}_2$
automorphism of the spin-4/3 fractional superconformal algebra
is realized in specific conformal field theory (CFT) models.

In Section \ref{sIII} we show how scattering amplitudes containing one
channel of either $R$ or $R'$ twisted-sector states and satisfying
spurious state decoupling can be constructed.  Various choices of
physical state conditions in the $R'$ sector are consistent with
spurious state decoupling in the prescription for scattering amplitudes
that we develop.  Presumably, demanding duality of four-point
amplitudes will further specify the set of $R'$ physical state
conditions.  The restriction in our scattering prescription to only two
external twisted-sector states is due to the absence of an appropriate
dimension-one vertex describing the emission of a single twisted-sector
state.  This is closely analogous to the situation in the ``old
covariant'' formalism of superstring amplitudes \cite{NST} in which,
due to the absence of the Faddeev-Popov superghost fields, an
appropriate dimension-one vertex for the emission of a Ramond-sector
state is also lacking.

This analogy is made even closer upon consideration of the example of
the $c=5$ model of the spin-4/3 string where the $R'$ sector is found
to describe space-time fermions.  In this model, discussed in Section
\ref{sIV}, the twisted-sector states are realized by a ${\bf Z}_2$
orbifold twisted sector of the $c=5$ CFT.  In particular, the $c=5$ CFT
is a tensor product of three free coordinate boson fields $X^\mu$ on
the world-sheet which do not participate in the orbifolding, and an
``internal'' $so(2,1)_2$ Wess-Zumino-Witten model described by two
bosons compactified on a triangular lattice.  The relevant
twisted sector arises upon twisting by a reflection through the
origin of that lattice.  Some details of this construction are
relegated to an Appendix.  In Section \ref{sIV} we also calculate some
of the low-lying physical states in the $R'$ sector and compute a
three-point coupling between two fermions and the massless vector boson
of the untwisted sector.  These computations place some restrictions on
the choice of physical state conditions; however, the precise set of
$R'$-sector physical state conditions consistent with duality of
tree amplitudes remains an open question in this model.

There are many other ways of twisting the triangular $so(2,1)_2$
lattice, but Lorentz invariance is invariably broken in the resulting
twisted sectors.   This follows from the fact that under the action of
all twists other than the ${\bf Z}_2$ twist used to define the $R'$
sector, at least one of the $so(2,1)_2$ currents is not invariant.
Currents not invariant under the twist can not have zero modes in the
twisted sector, implying that Lorentz invariance is broken in these
sectors.  Whether or not the states in various twisted sectors should
be included in the string spectrum is determined by one loop modular
invariance; however, since the $c=5$ model described above is not a
critical ($c=10$) representation of the spin-4/3 fractional
superstring, it is not expected to be modular invariant in any case.

In Section \ref{sV} we briefly describe a $c=2$ model of the spin-4/3
fractional superstring in which the $R$ (as opposed to $R'$) twisted
sector is realized.  Unlike the $R'$ sector, the $R$-sector physical
state conditions are determined by the scattering amplitude
prescription developed in Section \ref{sIII}.  Because the $c=2$ model
has only one flat dimension the resulting space-time physics
is trivial.  It remains an open question whether either the $R$ or $R'$
sector is actually realized in a critical ($c=10$) model of the
spin-4/3 fractional superstring.

\section{Twisted-sector representation theory}
\label{sII}

We begin with a brief description of the spin-4/3 fractional
superconformal (FSC) algebra.  A more detailed discussion can be found
in Ref.\ \cite{T1}.  The fractional currents, $G^\pm(z)$, and the
energy-momentum tensor, $T(z)$, together generate the FSC chiral
algebra, encoded in the singular terms of the operator product
expansions (OPE) \cite{ZFft}:
\widetext
	\begin{eqnarray}
	\label{fsca}
	T(z)T(w)&=&{1\over(z-w)^4}\left\{{c\over2}+2(z-w)^2T(w)
	+(z-w)^3\partial T(w)+\ldots\right\}~,\nonumber\\
	T(z)G^\pm(w)&=&{1\over(z-w)^2}\left\{{4\over3}G^\pm(w)
	+(z-w)\partial G^\pm(w)+\ldots\right\}~,\nonumber\\
	G^\pm(z) G^\pm(w)&=& {\lambda^\pm\over(z-w)^{4/3}} \left\{
	G^\mp(w)+{1\over2}(z-w)\partial G^\mp(w)+\ldots\right\},\nonumber\\
	G^\pm(z) G^\mp(w)&=& {1\over(z-w)^{8/3}} \left\{{3c\over8}
	+ (z-w)^2 T(w)+\ldots\right\}.
	\end{eqnarray}
\narrowtext \noindent
The first OPE implies that $T(z)$ obeys the conformal algebra with
central charge $c$, while the second implies that $G^\pm(z)$ are
dimension-4/3 Virasoro primary fields.  The constants $\lambda^\pm$ in
the $G^\pm G^\pm$ OPEs are fixed by associativity to be
	\begin{eqnarray}\label{lamb} \lambda^+ &=&
	\lambda^- = \sqrt{8-c\over6} ,\qquad{\rm for}\ c<8 ,\nonumber\\
	\lambda^+ &=& -\lambda^- =
	\sqrt{c-8\over6} ,\qquad{\rm for}\ c>8 .  \end{eqnarray}
This algebra generates the physical state conditions for the spin-4/3
fractional superstring.  Since there is only a single cut on the
right-hand side of each OPE, the currents $G^\pm$ are abelianly braided
(or parafermionic).  Under interchange of $z$ and $w$ (along a
prescribed path, say a counterclockwise switch) the only consistent
phase that $G^+$ or $G^-$ can pick up with itself is $e^{2i\pi/3}$.
The phase that develops upon interchange of $G^+$ with $G^-$ can be
taken to be $e^{-2i\pi/3}$.

The group of automorphisms of the FSC algebra organizes the
representation theory of its highest-weight modules.  The order-six
automorphism group $S_3$ of the FSC algebra is generated by the
transformations
\begin{mathletters}
	\label{autom}
	\begin{eqnarray}
	G^\pm &\rightarrow & \omega^{\pm1} G^\pm ,\\
	G^\pm &\rightarrow  & \delta G^\mp ,
	\end{eqnarray}
\end{mathletters} \noindent
where $\omega = e^{2\pi i/3}$, and $\delta = {\rm
sign}(8-c)$.  Since the FSC algebra is supposed to be an organizing
symmetry of the states of the spin-4/3 string, it is natural to assume
that its automorphisms extend to automorphisms of the CFT
representation of the FSC algebra.  All states can then be assigned
definite ${\bf Z}_3$ quantum numbers under the action of the ${\bf
Z}_3$ group of automorphisms generated by the transformation
(\ref{autom}a).  The untwisted sectors of the FSC algebra consist of
the set of states which obey the bypass relations
	\begin{equation}\label{untwistbypass}
	\chi_p(z) * \chi_q(w) = \omega^{2pq} \chi_p(z)\,\chi_q(w) ,
	\end{equation}
where $\chi_p$ is a state with ${\bf Z}_3$-charge $p$.  The bypass
relation $V(z) * W(w)$ denotes the analytic continuation of $z$ along a
closed path looping once around $w$ in a counterclockwise sense as
shown in Fig.\ \ref{fig1}a.  Clearly the currents themselves are
untwisted-sector fields with ${\bf Z}_3$ charges $\pm1$ for $G^\pm$
and charge 0 for $T$.

We define the twisted sectors of FSC algebra representations in terms
of the bypass relations the twist fields obey with the FSC currents.
The basic property of the twisted sectors is that the twist fields
are double-valued with respect to the fractional currents $G^\pm(z)$.
More precisely, with any twist field $\tau(w)$, the split
algebra currents satisfy the bypass relations
	\begin{eqnarray}\label{dblbypass}
	T * \tau &=& T\, \tau ,\nonumber\\
	G^\pm *^2 \tau &=& \omega G^\pm\,  \tau .
	\end{eqnarray}
The path defining the double-bypass relation $V(z) *^2 W(w)$ is shown
in Fig.\ \ref{fig1}b.  The phase in the double-bypass relation is
chosen so that the $G^\pm$ currents have zero modes in the twisted
sectors.  Indeed, the double-bypass relation implies that $z^{4/3}
G^\pm(z) \tau(0)$ is a double-valued analytic function on the $z$-plane
with branch point at $z=0$ and thus defines the mode expansion
	\begin{equation}\label{twistedom}
	 G^{\pm}(z)\tau(0)=\sum_{n\in{\bf Z}}z^{n/2-4/3}
	 G^{\pm}_{n/2}\tau(0).
	\end{equation}

To further specify the twisted sectors we must also know the single
bypass relations of the fractional currents $G^\pm$ with the twist
fields.  In general under a single bypass the $G^\pm$ currents go
to some other pair of (dimension-4/3) currents in the CFT under
consideration.  Denoting these currents by $\widetilde G^\pm$, we have
	\begin{eqnarray}\label{twcurbp}
	G^\pm * \tau &=& \delta\omega^2 \widetilde G^\mp \,\tau  ,
	\nonumber\\
	\widetilde G^\pm * \tau &=& \delta \omega^2 G^\mp\,\tau .
	\end{eqnarray}
We have included the factor of $\delta={\rm sign}(8-c)$ for later
convenience.  For these bypass relations to be consistent with
associativity of the CFT operator product algebra, $\delta \widetilde
G^\mp$ must be images of the $G^\pm$ currents under the action of some
${\bf Z}_2$ automorphism of the CFT.  This implies, in particular,
that $\widetilde G^\pm$ satisfy the same double-bypass relation
(\ref{dblbypass}) as $G^\pm$ do, implying that $\widetilde G^\pm$ have
half-integral mode expansions as in (\ref{twistedom}).  The single
bypass relations (\ref{twcurbp}) then imply the identification of modes
	\begin{equation}\label{ggtequiv}
	\widetilde G^\pm_{n/2} = \delta (-1)^n G^\mp_{n/2} .
	\end{equation}

There are some special choices for the $\widetilde G^\pm$ currents
(or, equivalently, for the ${\bf Z}_2$ automorphism) which lead
to particularly simple twisted-sector properties.  The first is
the choice $\widetilde G^\pm = G^\mp$, where the ${\bf Z}_2$
automorphism group is trivially realized by the identity
transformation.  In this case we simply recover the untwisted
sectors.  A second choice is $\widetilde G^\pm = \omega^{\mp p}
G^\pm$, where $p \in {\bf Z}_3$.  This choice
gives rise to twisted sectors we refer to as the $R$ sectors
of the FSC algebra, since they are equivalent to the $R$
sectors introduced in Ref.\ \cite{ZFft}.  The general case
where $\widetilde G^\pm \neq G^\pm$ gives rise to what we
call $R'$ twisted sectors.

The remainder of this section explores the representation theory of the
$R$ and $R'$ sectors.  We begin with the $R$ sector since it is
simpler, and derive the generalized commutation relations satisfied by
the modes of $G^\pm$ when acting on an arbitrary twisted-sector state.
We then turn to the $R'$ sector which is complicated by
the need to specify the operator product algebra of the $G^\pm$
currents with their images $\widetilde G^\pm$ under the ${\bf Z}_2$
automorphism. This depends, in general, on specific properties of the
CFT representaton of the FSC algebra in question.

\subsection{$R$-sector mode algebra}

The $R$ sectors arise from choosing the ${\bf Z}_2$ automorphism which
relates $G^\pm$ and $\widetilde G^\pm$ to be an automorphism of the FSC
algebra itself.  There are three ${\bf Z}_2$ subgroups of the $S_3$
automorphism group generated by the transformations (\ref{autom}) of
the spin-4/3 FSC algebra.  They give rise to three twisted sectors with
bypass relations
	\begin{equation}\label{Rbypass}
	G^\pm * \tau_p = \delta\omega^{2\pm p}G^\mp\,\tau_p ,
	\end{equation}
where $p\in{\bf Z}_3$.
(The $R$-sector
``$C$-disorder'' fields $\varphi_p$ introduced in Ref.\ \cite{ZFft}
actually satisfy the bypass relations $G^\pm * \varphi_p =
\delta\omega^{\pm p}G^\mp\,\varphi_{p\mp1}$.  With the change of basis
$\tau_p=\omega^2\varphi_p+\varphi_{p-1}+\varphi_{p+1}$ these bypass
relations become the bypass relations written above.) The single bypass
relations (\ref{Rbypass}) imply the mode identifications
	\begin{equation}\label{ZFmodiden}
	G^-_{n/2}=\delta \omega^{-p}(-1)^nG^+_{n/2} ,
	\end{equation}
when acting on $\tau_p$.  Thus in the $R$ sector there is really only
one independent fractional current, which we can take to be either
$G^+$ or $G^-$.

The three $R$ sectors labelled by $p\in {\bf Z}_3$ are related by the
${\bf Z}_3$ automorphism (\ref{autom}a) of the FSC algebra, and thus
are isomorphic in models in which that automorphism extends to the
whole CFT operator product algebra.  Since we assume that this is
always the case, we henceforth restrict the discussion to the $p=0$ $R$
sectors, the other two sectors being identical.

In analogy to the superconformal gauge of the ordinary superstring, the
physical states of the spin-4/3 fractional superstring are certain
highest-weight states of the FSC algebra---that is, they are
annihilated by all the positive modes of $T$ and $G^\pm$.  In order to
show spurious state decoupling in scattering amplitudes, we will need
to know the algebra satisfied by these modes.  This algebra takes the
form of generalized commutation relations (GCRs) for the modes of
$G^\pm$ due to cuts in the FSC operator product algebra
\cite{ZFpf,ZFft}.

To derive the $R$-sector GCRs, consider the integral
	\begin{eqnarray}\label{twistint}
	{\cal I} &=& {1\over4}\oint_\gamma {dz\over2\pi i}
	\oint_\delta{dw\over2\pi i}
	\left({\sqrt z+\sqrt w\over\sqrt z-\sqrt w}\right)^{\case{1}{3}}
	z^{\case{1}{3}+\case{n}{2}} w^{\case{1}{3}+\case{m}{2}}
	G^+(z)G^-(w)\tau(0) ,
	\end{eqnarray}
where the contours both wind twice around the origin with the $\delta$
contour inside the $\gamma$ contour.  The factors in the integrand have
been chosen to make these contours closed: the whole integrand is a
double-valued analytic function on both the $z$ and $w$ planes with
branch points at $z=0$ and $w=0$ and possible poles at $z=w$ or
$z=e^{2\pi i}w$ ({\it i.e.}, at the point $z=w$ on both sheets).
Evaluating this integral by shrinking the $\delta$ contour close to the
origin and using the mode definition (\ref{twistedom}), which can be
inverted as
	\begin{equation}
	 G^{\pm}_{n/2}\tau(0)={1\over2}{\oint_\gamma}
	 {{\rm d}z\over2\pi i}~z^{1/3+n/2}G^{\pm}(z)\tau(0),
	\end{equation}
gives
	\begin{equation}
	{\cal I}=\sum_{\ell=0}^\infty D^{(-{1\over3},{1\over3})}_\ell
	G^+_{(n-\ell)/2}G^-_{(m+\ell)/2}
	\end{equation}
where the $D^{(\alpha,\beta)}_\ell$ are binomial coefficients defined
by the expansion
	\begin{eqnarray}
	\label{Ddefn}
	 (1-x)^\alpha (1+x)^\beta=\sum_{\ell=0}^{\infty}
	 D^{(\alpha,\beta)}_\ell x^{\ell}.
	\end{eqnarray}

The integral ${\cal I}$ can also be evaluated in another way by
deforming the $\gamma$ contour to lie inside the $\delta$ contour.  The
result of this deformation is three contributions, as shown in
Fig.\ \ref{fig2}.  One contribution is just the same integral with
$\gamma$ and $\delta$ interchanged.  The other two contributions pick
up residues associated with the $G^+(z)G^-(w)\tau(0)$ OPE singularities
at $z=w$ and $z=e^{2\pi i}w$.  The former singularity can simply be
read off from the $G^+ G^-$ OPE; to evaluate the latter, we must
continue $z$ once counter-clockwise around the origin to $e^{2\pi i}z$
before letting it approach $w$ (on the second sheet).  By the
single-bypass relation (\ref{Rbypass}), this analytic continuation is
	\begin{equation}
	G^+(e^{2\pi i}z) G^-(w) \tau(0) =
	\delta \omega^2 G^-(z) G^-(w) \tau(0) .
	\end{equation}
Performing the same continuation on the other factors in the integrand,
picking up the residues of poles from the FSC algebra, and combining
all the contributions as shown in Fig.\ \ref{fig2}, gives
\widetext
	\begin{eqnarray}
	\label{RsectGCR}
	&& \delta\sum_{\ell=0}^\infty D^{({1\over3},-{1\over3})}_\ell
	\left[ G^+_{n-\ell\over2}G^+_{m+\ell\over2}+
	G^+_{m-\ell\over2}G^+_{n+\ell\over2}\right]
	= 2^{-5/3}\lambda^+ (-1)^{n+m}G^+_{n+m\over2}
	\nonumber\\ &&
	\qquad\qquad\mbox{}
	+2^{-4/3} \left\{(-1)^n+(-1)^m\right\}
	\left[L_{n+m\over2} +{c\over128}
	\left(6n^2-5\right)\delta_{n+m}\right].
	\end{eqnarray}
\narrowtext \noindent
where we have used (\ref{ZFmodiden}) to write the GCRs in terms of
$G^+$ modes alone.

The commutation relations of the $G^\pm$ modes with the $L_n$ modes
of the stress-energy tensor follow in the standard way:
	\begin{equation}\label{GLcom}
	\left[L_n,G^\pm_{m\over2}\right] =
	\left({n\over 3}-{m\over 2}\right)
	G^\pm_{n+{m\over2}}.
	\end{equation}
Acting on highest-weight states (states annihilated by all the positive
modes of $G^\pm$ and $T$), one can show from (\ref{RsectGCR}) that the
zero modes satisfy the relation
	\begin{equation}
	\label{Rsectzmalg}
	\delta\left(G^+_0\right)^2=2^{-8/3}\lambda^+ G^+_0
	+2^{-4/3} \left[L_0-{5c\over128}\right] .
	\end{equation}

\subsection{$R'$-sector mode algebra}

We refer to a twisted sector obeying the bypass relations
(\ref{twcurbp}) with $G^\pm \neq \widetilde G^\pm$ as an $R'$ sector.
The $R'$ sector of the FSC algebra is characterized by a ${\bf Z}_2$
automorphism of the particular CFT model in question which interchanges
$G^\pm\leftrightarrow \delta \widetilde G^\pm$.  This symmetry between
$G^\pm$ and $\delta \widetilde G^\pm$ implies that the $\widetilde
G\widetilde G$ OPEs form a second spin-4/3 FSC algebra by themselves
(at the same central charge and with structure constants
$\delta\lambda^\pm$), and that $\widetilde G^\pm$ have untwisted-sector
${\bf Z}_3$ charges $\pm1$, and so obey all the concomitant bypass
relations with $G^\pm$.  The $G\widetilde G$ OPEs, on the other hand,
remain undetermined by this symmetry,  and depend on the properties of
the specific CFT representation of the FSC algebra under
consideration.  We will assume that the $G \widetilde G$ OPEs are of
the form
	\begin{eqnarray}\label{GGtope}
	G^\pm(z)\widetilde G^\pm(w) &=& (z-w)^{-4/3} \left\{
	{\cal A}^\mp + \ldots\right\} ,\nonumber\\
	G^\pm(z)\widetilde G^\mp(w) &=& (z-w)^{-8/3} \delta \Biggl\{
	{3c\mu\over8}
	+ (z-w)^2 [\mu T+{\cal B}^\mp] + \ldots\Biggr\}.
	\end{eqnarray}
We will see that the $R'$ sector of the $c=5$ model to be considered
in Section \ref{sIV} has currents obeying OPEs of this form.

The form of the $G\widetilde G$ OPEs (\ref{GGtope}) is actually more
general, and follows for representations satisfying a few physical
properties.  Assume that these representations have a global
$D$-dimensional Poincar\'{e} symmetry realized by the tensor product of
$D$ coordinate bosons $X^\mu$ with an ``internal'' CFT which has a
positive definite spectrum of highest weights.  $G^\pm$ and $\widetilde
G^\pm$ then only involve derivatives of the $X^\mu$ fields with no
vertex contributions of the form $e^{ik\cdot X}$, and are
$so(D\!-\!1,1)$ singlets.  This requirement is natural in
representations which have a flat $D$-dimensional space-time
interpretation in string theory.  In addition, if no dimension-1/3 and
no dimension-one $so(D\!-\!1,1)$ scalar fields exist, then
(\ref{GGtope}) is the most general form the $G\widetilde G$ OPEs can
take.  Note that these last two conditions are not as strong as they
might appear.  For example, a pair of ${\bf Z}_3$-charge $\pm1$
dimension-1/3 scalar fields would obey braiding properties and operator
product selection rules identical to those of ${\bf Z}_3$ parafermion
currents; but associativity of the ${\bf Z}_3$ parafermion current
algebra fixes the central charge to be $c=4/5$ \cite{ZFpf}.  A similar
argument applies to potential dimension-one currents---for example, a
simple associativity argument shows that adding a single dimension-one
current to the spin-4/3 FSC algebra fixes the central charge to be
$c=1$.  These arguments are not proofs, though, since associativity
constraints can be evaded by increasing the number of independent
dimension-1/3 or dimension-one currents.  In any case, however the
reader judges the reasonableness of the above assumptions, there are at
least two cases in which they are satisfied---namely the $c=5$ model to
be discussed in detail in Section \ref{sIV}, and a $c=7$ model
described in Appendix C of Ref.\ \cite{T1}.

It follows from (\ref{GGtope})
that the combinations of fields $\Omega^\pm \equiv \widetilde
G^\pm -\delta \mu G^\pm$ are dimension-4/3, ${\bf Z}_3$-charge $\pm1$
untwisted-sector highest-weight operators with respect to the FSC
algebra generated by $G^\pm$.  The properties of the untwisted-sector
highest-weight modules derived in Ref.\ \cite{T1}, together with the
assumed symmetry under interchanging $G^\pm$ and $\delta\widetilde
G^\pm$,  implies that
	\begin{eqnarray}\label{ohtwo}
	\mu &=& {1\over3c}\left[c-32 \pm 2\sqrt{(8-c)(32-c)}\right]
	,\nonumber\\
	{\cal A}^\pm &=& {\mu \lambda^\mp \over \mu + 1}
	\left(G^\pm + \delta \widetilde G^\pm \right) .
	\end{eqnarray}
The dimension-two descendents of the $\Omega^\pm$ fields, $G^\mp_{-2/3}
|w^\pm\rangle$, are the dimension-two ${\cal B}^\mp$ fields that enter in
the $G^\pm\widetilde G^\mp$ OPE.

{}From the bypass relations (\ref{twcurbp}) and the OPEs (\ref{GGtope}),
and using the mode identifications (\ref{ggtequiv}), one deduces
the GCRs (just as was done in the last subsection)
\widetext
	\begin{eqnarray}\label{twistGCR}
	&&\sum_{\ell=0}^\infty D^{(-{1\over3},{1\over3})}_\ell\left[
	G^+_{n-\ell\over2}G^-_{m+\ell\over2}+
	G^-_{m-\ell\over2}G^+_{n+\ell\over2}\right]
	\nonumber\\ && \qquad\qquad          	
	= {2^{-5/3}\mu \over \mu+1}
	\left[(-1)^n \lambda^- G^+_{n+m\over2}+
	(-1)^m \lambda^+ G^-_{n+m\over2}\right]\nonumber\\ &&
	\qquad\qquad \ \mbox{}+2^{-4/3}\left\{1+(-1)^{n+m}\right\}
	\left[ L_{n+m\over2} + {c\over128}\left(6n^2-5\right)
	\delta_{n+m}\right],
	\end{eqnarray}
and
	\begin{eqnarray}\label{twistGCR3}
	&&\sum_{\ell=0}^\infty D^{(-{2\over3},{2\over3})}_\ell\left[
	G^\pm_{n-\ell\over2}G^\pm_{m+\ell\over2}-
	G^\pm_{m-\ell\over2}G^\pm_{n+\ell\over2}\right]
	=  2^{-5/3}\lambda^\pm(n-m) G^\mp_{n+m\over2}
	\nonumber\\ &&\qquad\qquad \mbox{} 	
	+ 2^{-4/3}
	\left\{(-1)^n+(-1)^m\right\}{3\mu c\over32}n\delta_{n+m}.
	\end{eqnarray}
The commutation relations (\ref{GLcom}) with the $L_n$ modes follow in
the standard way.

Note that, when acting on a highest-weight state,
the second GCR gives no relation for the $G^\pm_0G^\pm_0$ zero mode
product.  One can obtain a relation for these zero
modes by changing the power of the
$(\sqrt z-\sqrt w)/(\sqrt z +\sqrt w)$ factor in the integrand of
(\ref{twistint}).  One finds in this way
	\begin{eqnarray}\label{twistGCR2}
	&&\sum_{\ell=0}^\infty D^{({1\over3},-{1\over3})}_\ell
	\left[ G^\pm_{n-\ell\over2}G^\pm_{m+\ell\over2}+
	G^\pm_{m-\ell\over2}G^\pm_{n+\ell\over2}\right]
	= 2^{-5/3}\lambda^\pm G^\mp_{n+m\over2} \nonumber\\ &&
	\qquad\ \mbox{} +2^{-4/3} \left\{(-1)^n+(-1)^m\right\}
	\left[\mu L_{n+m\over2} + {\cal B}^\mp_{n+m\over2}
	+ {\mu c\over128}\left(6n^2-5\right)\delta_{n+m}\right].
	\end{eqnarray}
\narrowtext \noindent
Note, however, the appearance of the modes of the
representation-dependent dimension-two operators ${\cal B}^\pm$ in
(\ref{twistGCR2}).

\section{Scattering amplitudes}
\label{sIII}

We now extend the prescription developed in Ref.\ \cite{T1}
for dual N-point tree amplitudes of untwisted-sector states
satisfying spurious state decoupling to include two twisted-sector
states.  In that prescription amplitudes could be written in either of
the equivalent ``pictures''
        \begin{eqnarray}
        \label{Asum}
	{\cal A}_N &=& 2 \langle W^+|V(1) {1\over L_0-{1\over3}}
	V(1)\ldots{1\over L_0-{1\over3}} V(1)|W^-\rangle \nonumber\\
        &=& \langle V|V(1) {1\over L_0-1} V(1)\ldots
	{1\over L_0-1} V(1)|V\rangle
        \end{eqnarray}
where $W^\pm$ are untwisted-sector physical states of ${\bf Z}_3$
charge $\pm1$ and dimension 1/3, and $V$ is a certain dimension-one
descendent of these fields.

By analogy to the Ramond sector of the ordinary superstring in the
``old covariant'' formalism \cite{NST}, we take the physical states
$|\tau\rangle$ of the twisted sectors to be annihilated by the positive
modes of the currents $G^\pm$ and $T$, and to be eigenstates of their
zero modes with ``intercepts'' $h_t$ and $\Lambda^\pm$:
	\begin{eqnarray}
	\label{pscR}
	L_0 |\tau\rangle = h_t |\tau\rangle ,\nonumber\\
	G^\pm_0 |\tau\rangle = \Lambda^\pm |\tau\rangle .
	\end{eqnarray}
The $L_0$ condition gives rise to the mass-shell condition for the
physical states---a Klein-Gordon equation for each independent
component of the physical state.  The $G^\pm_0$ conditions, on the
other hand, are linear in space-time derivatives (because the
fractional currents have the form $G^\pm \sim \epsilon^\pm
\cdot\partial X + \ldots$ for models with a flat space-time
interpretation), and so should give rise to a Dirac equation for the
components of the physical states.  This will indeed turn out to be the
case in the models discussed in Sections \ref{sIV} and \ref{sV}.  Of
course, the Dirac equations must be consistent with the Klein-Gordon
equations.  This is automatically ensured in the $R$-sector by the
relation (\ref{Rsectzmalg}) between $G^\pm_0$ and $L_0$; however, it is
not automatically satisfied in the $R'$-sector, and must be imposed as
a separate requirement.

We expect the form of tree amplitudes with one twisted-sector channel
to consist of ``in'' and ``out'' twisted-sector physical states
$|\tau\rangle$ sandwiching the vertex operators of physical
untwisted-sector states $W$ which are strung together with the Dirac
propagator:
	\begin{equation}\label{ApictR}
	{\cal A}_N=\langle \tau_N|W^\pm_{N-1}(1)S^\pm\ldots S^\pm
	W^\pm_2(1)|\tau_1\rangle\ ,
	\end{equation}
where
	\begin{equation}
	S^\pm={1\over G^\pm_0-\Lambda^\pm} .
	\end{equation}
The choice of ${\bf Z}_3$ charges of the vertices and propagators
({\it i.e.}, their plus or minus superscripts) will be shown to be
immaterial in the $R$ sector, and will have to be further specified for
the $R'$ sector.  To show spurious state decoupling for these
amplitudes, we must be able to transform them to another ``picture'',
similar to those in (\ref{Asum}), in which the dimension-one $V$-vertices
appear instead of the dimension-1/3 $W^\pm$ vertices.  We first
describe this picture-changing and the resulting spurious-state
decoupling theorem in the $R$-sector case and then move onto the more
complicated $R'$-sector case.

\subsection{$R$-sector scattering amplitudes}

We start with the {\it anzatz} for the $N$-point amplitude with one
$R$-sector channel
	\begin{equation}
	\label{ApictRan}
	{\cal A}_N = \langle \tau_N | W^+_{N-1}(1)
	S^+ \ldots S^+ W^+_2(1) | \tau_1 \rangle .
	\end{equation}
This form for the amplitude can be rewritten in another ``picture''
using the following commutator
	\begin{eqnarray}\label{twiscom}
	[G^\pm_{n/2},V(1)] &=& \left(L_0+{n\over2}-h_t\right) W^\pm(1)
	-W^\pm(1)\left(L_0-h_t\right) ,
	\end{eqnarray}
for any integer $n$, which follows from the representation theory of
the untwisted-sector module that $V$ belongs to.  (See, for example,
the derivation of Eq.\ (3.10) of Ref.\ \cite{T1}.)  In particular,
rewrite a special case of (\ref{twiscom}) as
	\begin{eqnarray}\label{spcase}
	W^+(1)(L_0-h_t) &=& (L_0-h_t)W^+(1)-
	(G^+_0-\Lambda^+)V(1)
	+V(1)(G^+_0-\Lambda^+) .
	\end{eqnarray}
Insert a factor of $1=(L_0-h_t)/(L_0-h_t)$ before any $S^+$ propagator
in (\ref{ApictRan}) and commute the $(L_0-h_t)$ factor in the numerator
to the left using (\ref{spcase}).  The first two terms on the
right-hand side give vanishing contributions since the $(L_0-h_t)$
factor can be continually commuted to the left using (\ref{spcase})
until it annihilates the $\langle\tau_N|$ physical state, while the
second term vanishes by the ``cancelled propagator'' argument.  Tree
amplitudes with cancelled propagators are holomorphic in the Mandelstam
invariant of the cancelled propagator channel, and thus vanish by
analyticity if the amplitudes have sufficiently soft high-energy
behavior, as string amplitudes do.  The factor of $(G^+_0-\Lambda^+)$
in the third term of (\ref{spcase}) cancels the $S^+$ propagator,
leaving behind a Klein-Gordon propagator $\Delta=(L_0-h_t)^{-1}$.
Applying this argument repeatedly gives
	\begin{equation}\label{Apict2R} {\cal A}_N=\langle \tau_N|
	V_{N-1}(1)\Delta\ldots \Delta V_3(1)\Delta
	W^+_2(1)|\tau_1\rangle .  \end{equation}

The fact that one of the space-time boson vertices is still a
highest-weight state $W^+$, and not one of the ${\bf Z}_3$-charge $q=0$
descendents $V$, is familiar from tree amplitudes with one fermion line
for the ordinary superstring in the old covariant formalism .  The
position of the lone $W^+$-vertex is arbitrary, as can be shown by
manipulations similar to those used to derive (\ref{Apict2R}).
Furthermore, the above manipulations can be reversed using
Eq.\ (\ref{spcase}) with $W^-$, $G^-_0$, and $\Lambda^-$ to show
that any choice of $\pm$ superscripts in (\ref{ApictR}) is equivalent
to (\ref{ApictRan}).  It is important to note in this connection that
due to the equivalence of $G^+$ and $G^-$ modes (\ref{ZFmodiden}) in
the $R$-sector, that we must have $\Lambda^+=\delta \Lambda^-$.

The following trick allows us to re-express the amplitude
(\ref{Apict2R}) completely in terms of $V$ vertices.  Insert
$1=(L_0-h_t)/(L_0-h_t)$ between $W^+_2(1)$ and $|\tau_1\rangle$ and use
(\ref{spcase}) once more to finally obtain the amplitude in the form
	\begin{equation}\label{Apict3R}
	{\cal A}_N=\langle
	\tau_N|V_{N-1}(1)\Delta\ldots\Delta V_2(1)
	{G^+_0-\Lambda^+\over L_0-h_t}|\tau_1\rangle .
	\end{equation}
This form for the amplitude is slightly delicate because both the
numerator and denominator in the factor before $|\tau_1\rangle$
annihilate it.  However, this potential ambiguity can be resolved by
noting that the $L_0-h_t$ operator, when acting on the spinor
wavefunction $u(k)$ associated with $\tau_1$, is the Klein-Gordon
operator $k^2+m^2$, while $G^+_0-\Lambda^+$ is the
associated Dirac operator $ik\cdot\gamma +m$.  For massive states the
simple zeroes of the numerator and denominator cancel, and for massless
states it can be defined as the limit as $k$ goes on mass-shell.

If we start with physical states in the amplitude (\ref{Apict3R}), will
they scatter only to other physical states?  We can reformulate this
question in terms of spurious state decoupling.  A state $\langle s|$
obeying the zero-mode conditions in Eq.\ (\ref{pscR}) is called a
spurious state if it is orthogonal to all physical states.  Since the
physical state conditions are the only restriction on a
generic physical state, it follows that $\langle s|$ can be written as
	\begin{equation} \label{spur}
	\langle s| = \sum_{m>0} \langle \chi_m| L_m +
	\sum_{n>0} \langle \psi^\pm_{n\over2}| G^\pm_{n\over2}
	\end{equation}
in terms of some other states $\chi_m$ and $\psi^\pm_{n\over2}$.  All
states not satisfying the physical state conditions must have a
spurious component.  A physical state can itself be spurious, in which
case it is a null state (since it is orthogonal to itself), and should
also decouple from all scattering amplitudes.  Thus, the decoupling of
all spurious states from scattering amplitudes of physical states is a
prerequisite for a sensible interpretation of those amplitudes.  For
this decoupling to be true, no spurious states should contribute to
residues of poles in amplitudes when an internal propagator goes
on-shell.

Suppose we fix the external momenta $k_i$ of the vertices $V_i$ such
that some state $|s\rangle$ in the string Fock space at momentum $\kappa
=k_{M+1}+\cdots+k_N$ is on-shell:  $(L_0-h_t)|s\rangle=(G^\pm-\Lambda^\pm)
|s\rangle=0$.  Factorize the amplitude in Eq.\ (\ref{Apict3R}) by
inserting a sum over a complete set of states $\sigma_i$ of momentum
$\kappa$ at the propagator between $V_{M+1}$ and $V_M$,
\widetext
	\begin{eqnarray}
	{\cal A}_N &=& \sum_i \langle \tau_N |
	V_{N-1}(1) \ldots \Delta V_{M+1}(1) {G^+_0-\Lambda^+\over
	L_0-h_t} |\sigma_i\rangle {1\over i\kappa\cdot\gamma+m_i}
	\nonumber\\ &&\qquad\qquad \times	
	\langle\sigma_i| V_M(1) \ldots \Delta V_2(1)
	{G^+_0-\Lambda^+\over L_0-h_t} |\tau_1\rangle .
	\end{eqnarray}
\narrowtext \noindent
Then the $|s\rangle\langle s|$ term in the sum contributes a pole
in momentum space.  The requirement of spurious state decoupling is
that if $|s\rangle$ is spurious, its contribution to the residue of the
pole should vanish:
	\begin{equation}
	\label{spstde}
	\langle s|V_M(1)\Delta\cdots\Delta V_2(1)
	{G^+_0-\Lambda^+\over L_0-h_t}|\tau_1\rangle=0 .
	\end{equation}

To prove this, consider one term, say $\langle\psi|G^+_{n/2}$ with
$n\ge0$, in the presentation of $\langle s|$ as a sum of descendent
states, Eq.\ (\ref{spur}), where $\psi$ must satisfy
$(L_0+{n\over2}-h_t) |\psi\rangle = 0$.  (The $G^-_{n/2}$ descendent
pieces can be shown to decouple by the same argument using the
$R$-sector equivalence of $G^+$ and $G^-$ (\ref{ZFmodiden}).  The $L_n$
pieces decouple by a simpler argument.) The $G^+_{n/2}$ mode can be
commuted to the right in Eq.\ (\ref{spstde}) using Eq.\ (\ref{twiscom})
and the identity
	\begin{equation}
	G^\pm_{n\over2} {1\over L_0-h_t} =
	{1\over L_0-h_t-{n\over2}} G^\pm_{n\over2} ,
	\end{equation}
which follows from (\ref{GLcom}).  The insertions coming from the
right-hand side of Eq.\ (\ref{twiscom}) again vanish by a cancelled
propagator argument.  Finally, the $G^+_{n/2}$ mode can be seen to pass
through the $G^+_0-\Lambda^+$ factor and annihilate $|\tau_1\rangle$ by
the physical state conditions using the $R$-sector GCR
(\ref{RsectGCR}).  This proves spurious state decoupling.

Since there is no appropriate dimension-one commuting vertex in the
$R$ sector, we cannot extend our scattering amplitude prescription
(\ref{ApictRan}) or (\ref{Apict2R}) to include more than two $R$-sector
vertices.  By the same token, we cannot prove cyclic symmetry
(duality) of these amplitudes in the present formalism.
This situation is closely analogous to what happens in the old
covariant formalism in the ordinary superstring.  There, dual
amplitudes with spurious state decoupling can be formulated for
scattering of Neveu-Schwarz sector states, and can only be extended to
include two Ramond-sector states as the ``in'' and ``out'' states in
the correlator, thus losing manifest cyclic symmetry.  So presumably,
just as in the Ramond sector of the superstring, our inability to
incorporate more than two $R$-module physical states in our scattering
prescription means that there is a nontrivial contribution to $R$-module
scattering amplitudes coming from the ``fractional superghost'' fields
on the world-sheet.

\subsection{$R'$-sector scattering amplitudes}

The main difference between the $R$ and $R'$ sector scattering
amplitudes is that in the $R'$ sector there is no identification of
$G^+$ with $G^-$ modes analogous to (\ref{ZFmodiden}).  This means that
the zero-mode conditions (\ref{pscR}) generate two {\it a
priori\/} independent Dirac equations for the physical state wave
functions.  This will lead to overly restrictive physical
state conditions unless the $G^\pm$ intercepts are appropriately tuned,
or only a subset of all the $G^\pm_{n/2}$ modes with $n\ge0$ are used
as $R'$-sector physical state conditions.  Since, as was emphasized at
the end of Section \ref{sII}, the mode algebra obeyed by the fractional
currents in the $R'$ sector depends on the details of the particular
CFT model under consideration, we can only hope to determine the
appropriate $R'$-sector physical state conditions in the context of
a specific model.

However, we can still demonstrate spurious state decoupling in
amplitudes with one $R'$-sector channel.  Indeed, all the manipulations
of the last subsection for the $R$ sector scattering go through
unchanged in the $R'$ sector as long as all the untwisted vertices
$W^\pm$ and Dirac propagators $S^\pm$ in (\ref{ApictR}) have the same
${\bf Z}_3$ charge ({\it i.e.}, all their superscripts are the same).
The point is simply that since the $G^+_0$ and $G^-_0$ modes are not
related in any general way in the $R'$ sector, the associated Dirac
propagators are not equivalent---in particular $(G^+_0 - \Lambda^+
)/(G^-_0 - \Lambda^-)$ is not proportional to the identity.

In fact, one can show that the {\it anzatz} for the one channel $R'$-sector
scattering amplitude
	\begin{equation}\label{ApictRprime}
	{\cal A}_N=\langle \tau_N|W^\alpha_{N-1}(1)S^\alpha\ldots
	S^\alpha W^\alpha_2(1)|\tau_1\rangle
	\end{equation}
obeys spurious state decoupling,
where $\alpha$ is any fixed complex number and we define
	\begin{eqnarray}
	G^\alpha_{n/2} &=& G^+_{n/2} + \alpha G^-_{n/2} ,\nonumber\\
	\Lambda^\alpha &=& \Lambda^+ + \alpha \Lambda^- ,\nonumber\\
	S^\alpha &=& \left(G^\alpha_0 -\Lambda^\alpha \right)^{-1},
	\nonumber\\
	W^\alpha &=& W^+ + \alpha W^- .
	\end{eqnarray}
This follows from the fact that the analog of the commutation
relation (\ref{twiscom}) is satisfied for the above combinations:
	\begin{eqnarray}
	[G^\alpha_{n/2},V(1)] &=&
	\left(L_0+{n\over2}-h_t\right) W^\alpha(1)
	-W^\alpha(1)\left(L_0-h_t\right) ,
	\end{eqnarray}
Thus, we have a one-parameter family of satisfactory scattering
amplitudes in the $R'$ sector.

Tree-level duality may restrict the value of $\alpha$ and the correct
set of physical state conditions in the $R'$ sector.  In particular, by
duality, if one factorizes the amplitude (\ref{ApictRprime}) in a
channel other than the $R'$-sector channel displayed, one should find
an infinite tower of poles corresponding to untwisted-sector physical
states.  Presumably this occurs only if the $R'$-sector states
$\tau$ obey the correct set of physical state conditions.  We will
discuss the choice of physical state conditions in the context of a
concrete example in the next section, though to the level we compute we
will not be able to put many restrictions on the possible choices.  As
an example of the kind of choices that could make sense, note that
associated with each value of $\alpha$ is a natural choice of
physical state conditions for $R'$-sector states:
	\begin{eqnarray}
	\label{pscRprime}
	\left(L_n-h_t\delta_{n,0}\right)|\tau\rangle &=& 0, \nonumber\\
	\left(G^\alpha_{n/2}-\Lambda^\alpha\delta_{n,0}\right)
	|\tau\rangle &=& 0,
	\end{eqnarray}
for $n\ge0$.  Note also that by the commutation relation (\ref{GLcom})
all the possible physical state conditions are generated by $L_0$,
$L_1$, $G^\pm_0$, and $G^\pm_{1/2}$.  In particular, the commutator of
$L_1$ with $G^\pm_0$ generates all the integer-moded $G^\pm_m$
conditions.  Thus, it may also be consistent to restrict the physical
state conditions to only the integer-moded $T$ and $G^\alpha$
annihilation operators.  These, however, are only the simplest
guesses.  The fact that $R'$-sector states are ${\bf Z}_2$-twist
fields obeying non-trivial monodromies with each other may indicate
that the physical state conditions and scattering amplitudes for
$R'$-sector ``matter'' fields ({\it i.e.}, after integrating out the
fractional superghost pieces of the full physical vertices) may be more
complicated than what we have presented here.

\section{Space-time fermions in the \lowercase{c}=5 model}
\label{sIV}

Strings propagating in $D$ flat space-time dimensions are described
by a world-sheet CFT which includes $D$ massless scalar fields
$X^\mu(z)$.  The spin-4/3 fractional superstring CFT also includes
a set of fields $\epsilon^\pm_\mu(z)$ of conformal dimension 1/3,
transforming as vectors under space-time Lorentz transformations. The
simplest non-trivial such CFT which is also a representation of the
spin-4/3 FSC algebra can be constructed from five free massless scalar
fields on the world-sheet, and hence has central charge $c=5$.  (A list
of known representations of the spin-4/3 FSC algebra is given in
Appendix C of Ref.\ \cite{T1}.) Three of the scalars are just
coordinate boson fields $X^\mu(z)$, $\mu=0,1,2$, with the standard
operator products $X^\mu(z) X^\nu(w) = -g^{\mu\nu}\ln(z-w)$, where
$g^{\mu\nu}$ is the three-dimensional Minkowski metric with signature
$(-++)$.  The remaining two fields, $\varphi^i(z)$, $i=1,2$, are
compactified on a triangular lattice:
$\varphi^i(z)=\varphi^i(z)+2\pi$, with $\varphi^i(z)\varphi^j(w) =
-g^{ij}\ln(z-w)$, where
        \begin{equation}
        g^{ij}={1\over3}\pmatrix{\hfill 2&-1\cr -1&\hfill 2\cr} .
        \end{equation}
These two bosons form a representation of the $so(2,1)_2$
Wess-Zumino-Witten model.  Thus the $c=5$ model has a global
three-dimensional Poincar\'e symmetry.

Vertex operators in the $so(2,1)_2$ CFT, $V_{\bf m} = {\rm exp}
\{im_1\varphi^1 + im_2\varphi^2\}$ for integer $m_i$, describe
Virasoro-primary operators transforming as integer-spin fields under
the $so(2,1)$ symmetry.  For example, some of the simplest $so(2,1)_2$
Virasoro-primary fields are the dimension-1/3 $so(2,1)$-vector fields
$\epsilon^\pm_\mu$, the dimension-4/3 $so(2,1)$-scalars $s^\pm$, and
the dimension-one $U_\mu$'s, given by
        \begin{eqnarray}
   	\label{covflds}
	\epsilon^\pm_\mu &=& \left(V_{(\mp1,\mp1)}\,,\, V_{(\pm1,0)}\,,\,
	V_{(0,\pm1)} \right) \nonumber\\
	s^\pm &=& \case{1}{3} \left( V_{(\pm2,\pm2)}\!+\!V_{(\mp2,0)}\!+\!
	V_{(0,\mp2)} \right) \nonumber\\
	U_\mu &=& \Bigl(V_{(1,-1)}\!+\!V_{(-1,1)}\,,\,
	V_{(1,2)}\!+\!V_{(-1,-2)}\,,\,
	V_{(2,1)}\!+\!V_{(-2,-1)} \Bigr) .
        \end{eqnarray}
The properties of these and other fields in the $c=5$ model are
discussed in more detail in Ref.\ \cite{T1}.

The fractional supercurrents $G^\pm$ satisfying the spin-4/3 FSC
algebra (\ref{fsca}) are given in terms of these fields by
        \begin{equation}
	\label{Grepd}
	G^\pm = {1\over\sqrt2} \left(\pm\epsilon^\pm\cdot\partial X
        -{3\over2}s^\pm \right) .
        \end{equation}
There are actually six solutions for the supercurrents, which can be
obtained from the above solution by making the transformations $G^\pm
\rightarrow \omega^{\pm q} G^\pm$ or $G^\pm \rightarrow \omega^{\pm q}
\widetilde G^\pm$, where $q\in{\bf Z}_3$ and $\widetilde G^\pm$ are
given by
        \begin{equation}
	\label{Grepspltil}
	\widetilde G^\pm = {1\over\sqrt2} \left(\mp\epsilon^\pm
	\cdot\partial X -{3\over2}s^\pm \right) ,
        \end{equation}
which differs from (\ref{Grepd}) by a sign change in the
$\epsilon^\pm\cdot\partial X$ terms.  The existence of these six
solutions is a consequence of the ${\bf Z}_2\times S_3$ automorphism
group of the $c=5$ CFT generated by $X^\mu \rightarrow -X^\mu$, $V_{\bf
m} \rightarrow \omega^{m_1+m_2} V_{\bf m}$, and $V_{\bf m} \rightarrow
V_{-{\bf m}}$, which leaves the $so(2,1)$ generators invariant.

In what follows, we construct the $R'$ twisted sector
corresponding to the ${\bf Z}_2$ automorphism $V_{\bf m} \rightarrow
V_{-{\bf m}}$ which is generated by the transformation $\varphi^j
\rightarrow -\varphi^j$.  Recalling the discussion of Section
\ref{sII}, the fields of this twisted sector are characterized by the
property that under single-bypass around a twisted-sector field,
untwisted-sector fields transform according to this automorphism.
Thus the defining property of the $R'$ twisted sector is that with any
twist field, $\tau(w)$, the free boson fields ${\varphi^j}(z)$ satisfy
the basic bypass relation
	\begin{equation}
	\label{Btwistby}
	{\varphi^j} * \tau = -{\varphi^j} \, \tau .
	\end{equation}
This is also the definition of the twisted sector of a ${\bf Z}_2$
orbifold \cite{orbs} of the $so(2,1)_2$ CFT under the action of the
symmetry which reflects the triangular $\varphi^j$-boson lattice
through the origin.  Note that this ${\bf Z}_2$ transformation maps
the $G^\pm$ currents into the $\widetilde G^\pm$ currents.  It is
straight forward to show that these currents obey OPEs of the form
(\ref{GGtope}) with
	\begin{eqnarray}\label{cfiveops}
	\mu &=& -3/5 , \nonumber\\
	{\cal A}^\pm &=& - {3\over2\sqrt2}
	(G^\pm + \widetilde G^\pm) , \nonumber\\
	{\cal B}^\pm &=& {3\over5} T_\varphi - {2\over5} T_X \pm
	{1\over2} U\cdot\partial X ,
	\end{eqnarray}
where $T_X$ and $T_\varphi$ are the stress-energy tensors for the
$X^\mu$ and $so(2,1)_2$ CFTs, respectively.  Thus we are indeed
constructing precisely the $R'$ sector discussed in Section \ref{sII}.

There is a physical reason for expecting this sector to appear in the
spin-4/3 fractional superstring: it includes the space-time fermionic
states of the $c=5$ model.  The untwisted-sector states described so
far are all space-time bosonic states, corresponding to integer-spin
representations of $so(2,1)$, in the $c=5$ FSC algebra representation.
In general, the $so(N)_2$ Wess-Zumino-Witten model can be realized as
the ${\bf Z}_2$ orbifold of the $su(N)_1$ model, with the spinor
representations of $so(N)$ appearing as the ${\bf Z}_2$-twisted fields
\cite{Itoh}.  Thus we expect to find the space-time fermionic physical
states of the $c=5$ model of the spin-4/3 fractional superstring in the
${\bf Z}_2$-twisted sector of $so(2,1)_2$.  In the next subsection we
construct this twisted sector in some detail, and show along the
way that the $R'$-sector bypass relations (\ref{twcurbp}) are indeed
satisfied.  In the subsequent subsections we compute some
low-lying $R'$-sector physical states and discuss their scattering
amplitudes using the results of Section \ref{sIII}.

\subsection{The twisted sector of $so(2,1)_2$}

{}From the bypass relation (\ref{Btwistby}), it follows that the
${\varphi^j}(z)$ bosons have the mode expansion
	\begin{equation}\label{Btwmodeexp}
	\varphi^j(z)=\phi^j+i\sum_{r\in{\bf Z}+{1\over2}}{1\over r}
	\beta^j_r z^{-r} , \end{equation}
with modes satisfying the commutation relations
	\begin{equation}\label{betacomm}
	[\beta^i_r,\beta^j_s]=r g^{ij}\delta_{r+s} .
	\end{equation}
It is easy to show that
	\begin{equation}
	\left[i\sum_{r>0}{1\over r}\beta^i_r z^{-r}\, ,\,
	i\sum_{s<0}{1\over s}\beta^j_s w^{-s}\right] = -g^{ij}{\rm
	ln}\left({\sqrt z-\sqrt w\over\sqrt z+\sqrt w}\right) ,
	\end{equation}
implying the basic OPE $\varphi^i(z) \varphi^j(w) = -g^{ij} {\rm
ln}(z-w) +\ldots$.  The quantization of the zero mode $\phi^j$ is crucial
for correctly reproducing the operator products of untwisted sector
operators when acting on states in the twisted sector.  As shown in
Ref.\ \cite{Itoh}, the correct quantization of the $\phi^j$ zero modes
of ${\varphi^j}(z)$ is
	\begin{equation}\label{narain}
	[\phi^i,\phi^j]=i\pi\varepsilon^{ij},
	\end{equation}
where $\varepsilon^{ij}$ is the antisymmetric two-index tensor
normalized by $\varepsilon^{12}=1$.  These commutation relations result
from the proper quantization of free boson zero modes in the presence
of the constraints arising from identifications under lattice
translations and the inclusion of a constant antisymmetric background
field \cite{Itoh}.

In the twisted sector ({\it i.e.}, with a ${\bf Z}_2$ twist field
inserted at the origin) the basic vertex operators $V_{\bf m} =
:e^{i{\bf m}\cdot{\bbox\varphi}}:$ have the normal ordered expression
\widetext
	\begin{eqnarray}\label{twvtx}
	V_{\bf m}(z) &=& {\rm exp}\left\{-{i\pi\over2}m_1m_2\right\}
	2^{-{\bf m\cdot m}} z^{-{1\over2}{\bf m\cdot m}}
	{\rm exp}\left\{i{\bf m}\cdot{\bbox \phi}\right\}
	\nonumber\\ &&\quad\times 	
	{\rm exp}\left\{-\sum_{r<0}{1\over r}
	{\bf m}\cdot\bbox{\beta}_r z^{-r}\right\}
	{\rm exp}\left\{-\sum_{r>0}{1\over r}
	{\bf m}\cdot\bbox{\beta}_r z^{-r}\right\} .
	\end{eqnarray}
\narrowtext \noindent
The first factor provides the signs needed to ``Wick rotate'' from
$so(3)$ to $so(2,1)$.  No further cocycles are needed, since the
$so(3)$ signs are automatically taken care of by the commutation
relations of the $\phi^j$ zero-modes.  The normal ordering factors
$(2\sqrt z)^{-{\bf m\cdot m}}$ are required to insure the
factorizability of amplitudes (associativity of the operator product).
Using this explicit form (\ref{twvtx}) for the vertex and the mode
commutation relations (\ref{betacomm}) and (\ref{narain}), it is easy
to check that the vertex operators indeed obey the correct
untwisted-sector operator product expansion
	\begin{equation}\label{Bbope}
	V_{\bf m}(z) V_{\bf n}(w) =
	(-1)^{m_2n_1}(z-w)^{\bf
	m\cdot n} V_{\bf m+n}(w) + \ldots
	\end{equation}
as they should since the operator products encode only local
information and do not depend on whether any twist fields are located
elsewhere on the world-sheet.

The representations of the zero mode algebra (\ref{narain}) determine
the properties of the twisted-sector ground state.  It is shown in
the Appendix that there are only three representations of the zero mode
algebra which preserve the global $so(2,1)$ symmetry of the $so(2,1)_2$
CFT.  This implies that there are three twisted sectors ${\cal T}_p$,
labeled by $p\in{\bf Z}_3$, and that if $\tau_p$ is an arbitrary twist
field in ${\cal T}_p$ and $V_{\bf m}$ is an untwisted-sector vertex
operator, then $V_{\bf m}\tau_p\in{\cal T}_p$.  Thus the three twisted
sectors are disjoint; in fact, they are just copies of one another
under the action of a ${\bf Z}_3$ symmetry of the $so(2,1)_2$ CFT.
Indeed, from the expression (\ref{twvtx}) for the $V_{\bf m}$ vertex
acting on an arbitrary twisted-sector state $\tau_p(0)$ one obtains the
bypass relation
	\begin{eqnarray}\label{sbpr}
	V_{\bf m}(z) * \tau_p(0)&=& e^{-i\pi{\bf m\cdot m}}
	e^{2i{\bf m}\cdot{\bbox \phi}} V_{-{\bf m}}(z)\, \tau_p(0)
	\nonumber\\ &=& e^{-i\pi{\bf m\cdot m}} \omega^{p(m_1+m_2)}
	V_{-{\bf m}}(z)\, \tau_p(0) , \end{eqnarray}
where in the second line we have used a property of the zero-mode
representations derived in the Appendix.  This bypass relation implies
the double-bypass relation $V_{\bf m} *^2 \tau_p = e^{-2i\pi{\bf m\cdot
m}}V_{\bf m}\,\tau_p$, which in turn implies the existence of the mode
expansion
	\begin{equation}\label{twvtxmodexp}
	V_{\bf m}(z)\tau_p(0)=\sum_{n\in{\bf Z}}
	z^{-{1\over2}{\bf m\cdot m} -{n\over2}} (V_{\bf
	m})_{n\over2}\tau_p(0) .  \end{equation}
It is clear from (\ref{sbpr}) that the ${\cal T}_p$ twisted sectors can
be ``rotated'' into each other by the ${\bf Z}_3$ automorphism of the
$so(2,1)_2$ CFT:  $V_{\bf m}\rightarrow \omega^{m_1+m_2} V_{\bf m}$.
Thus the three twisted sectors are equivalent and we henceforth
restrict ourselves (without loss of generality) to the $p=0$ sector.
In this sector the single-bypass relation (\ref{sbpr}) for the basic
fields are
	\begin{eqnarray}
	\epsilon^{\mu\pm} * \tau &=& \omega^2 \epsilon^{\mu\mp} \tau
	\nonumber \\
	s^\pm * \tau &=& \omega^2 s^\mp \tau \nonumber \\
	G^\pm * \tau &=& \omega^2 \widetilde G^\mp \tau ,
	\end{eqnarray}
the last of which is precisely the defining bypass relation
(\ref{twcurbp}) of the $R'$ sector.  These bypass relations along with
the mode expansion (\ref{twvtxmodexp}) imply that ${\bf Z}_2$-even
fields have only integer moding, while the ${\bf Z}_2$-odd fields have
only half-odd integer modings.  For example,
	\begin{eqnarray}
	\label{so21modiden}
	\epsilon^{\mu+}_{n/2} &=& (-1)^n \epsilon^{\mu-}_{n/2}
	\nonumber\\
	s^+_{n/2} &=& (-1)^n s^-_{n/2} \nonumber\\
	G^\pm_{n/2} &=& (-1)^n \widetilde G^\mp_{n/2} .
	\end{eqnarray}

We now discuss the properties of the twist fields.  In order to
characterize the twisted-sector states, we must take into account the
action of the zero modes $\phi^j$ on the twist states.  In particular,
the twisted-sector ground state forms a representation of the
nontrivial zero-mode algebra (\ref{narain}).  As a result, it is shown
in the Appendix, the ground states in the twisted sectors are doubly
degenerate.  We label the corresponding twist fields by $\sigma^a(z)$
and the ground states by $|a\rangle =\sigma^a(0)|0\rangle$,
where $a=0$ or $1$.

The action of the zero mode of an arbitrary untwisted-sector operator
on the twisted-sector ground states can be worked out using the methods
of the Appendix.  In particular, one finds that the zero modes of the
$so(2,1)$ scalar and vector fields act on the ground state as
	\begin{eqnarray}\label{gammazero}
	s^\pm_0|a\rangle &=& 2^{-8/3}|a\rangle ,\nonumber\\
	\epsilon^{\mu\pm}_0|a\rangle &=& 2^{-2/3}
	(\gamma^\mu)^a_b|b\rangle ,\nonumber\\
	U^\mu_0|a\rangle &=& 2^{-1} (\gamma^\mu)^a_b |b\rangle ,
	\end{eqnarray}
where $\gamma_\mu$ is a Dirac gamma matrix obeying
	\begin{equation}
	\label{gammalg}
	\gamma^\mu\gamma^\nu = g^{\mu\nu}
	-\varepsilon^{\mu\nu\rho}\gamma_\rho ,
	\end{equation}
where $\varepsilon^{\mu\nu\rho}$ is the antisymmetric tensor in three
dimensions normalized by $\varepsilon^{012}=1$.  These zero-mode
actions imply that the twisted sector ground states transform as
spinors under the global $so(2,1)$ symmetry.  Thus twisted sector
physical states describe space-time fermionic excitations of the
spin-4/3 fractional string.

Since the ground states satisfy $\beta^j_r|a\rangle = 0$ for
$r>0$, a simple computation reveals
\widetext
	\begin{equation}
	\label{dimcomp}
	\langle a| T(z) |b\rangle =
	\lim_{w\rightarrow z} \langle a| \left\{ -{1\over2}
	\partial\varphi^i(w) g_{ij} \partial\varphi^j(z) -
	{1\over(w-z)^2} \right\} |b\rangle =
	C_a^b{1\over8}z^{-2},
	\end{equation}
\narrowtext \noindent
implying that the conformal dimension of the twisted-sector ground
state is $h(\sigma^a)=1/8$.  $C_a^b$ is the spinor metric  which can be
taken to be $(\gamma^0)^b_a$.  All other twist fields are created from
$\sigma^a$ by the repeated action of the ${\varphi^j}$ creation modes
$\beta^j_r$ with $0>r\in{\bf Z}+{1\over2}$.  Thus the twist states
have the spectrum of conformal dimensions $h={1\over8}+{n\over2}$
for $n\ge0$ an integer.

\subsection{Mode algebra in the $so(2,1)_2$ twisted sector}

Acting on a twisted-sector state, the FSC currents $G^\pm$ have the
mode expansion
	\begin{equation}
	G^\pm(z)\tau(0) = \sum_{n\in{\bf Z}} z^{-{4\over3}-{n\over2}}
	G^\pm_{n/2} z^{-n/2} \tau(0) ,
	\end{equation}
following from (\ref{twvtxmodexp}).  Since the space-time coordinate
boson fields $X^\mu$ of the $c=5$ representation of the FSC algebra are
unaffected by the orbifolding procedure, their mode expansion is the
usual one,
	\begin{equation}
	\label{Xbosexp}
	X^\mu(z) = x^\mu - i\alpha_0^\mu {\rm ln}(z) +
	i\sum_{n\neq0} {1\over n} \alpha^\mu_n z^{-n}  ,
	\end{equation}
satisfying the standard commutation relations $[x^\mu, \alpha_0^\nu] =
ig^{\mu\nu}$ and $[\alpha^\mu_m, \alpha^\nu_n] = m\delta_{m+n}
g^{\mu\nu}$.  Combining this with the mode expansion of the $so(2,1)_2$
fields acting on twisted-sector states, and recalling the form of the
currents (\ref{Grepd}), we obtain
	\begin{equation}\label{Gtwmodexp}
	G^\pm_{n/2} = {1\over\sqrt2} \left\{ \mp i\sum_{m\in{\bf Z}}
	\alpha_{m} \cdot \epsilon^\pm_{{n\over2}-m} -
	\case{3}{2} s^\pm_{n\over2} \right\} .
	\end{equation}

Though we could, in principle, use the mode expansion (\ref{Gtwmodexp})
to calculate the action of the $G^\pm$ modes on arbitrary
twisted-sector states using the expression (\ref{twvtx}) for the vertex
operators in terms of the modes of $\varphi^j$, this is in practice a
complicated way of computing.  A more
efficient way which preserves Lorentz invariance in intermediate steps
is to express all the twisted-sector states  in terms of products of
modes of $\epsilon_\mu^\pm$ acting on the twisted-sector ground-state
spinor.  The set of all products of $\epsilon_\mu^\pm$ modes is not a
linearly independent basis, however, so we need to compute the
algebra satisfied by the $\epsilon_\mu^\pm$ modes.  Also, in order to
compute the action of the $G^\pm$ and $T$ modes on twisted-sector
states, we have to express them in terms of the $\epsilon_\mu^\pm$
modes.

The algebra of the $\epsilon_\mu^\pm$ modes can be written as a
set of generalized commutation relations (GCRs) derived from
the $\epsilon_\mu^\pm$ OPEs in the same way that the $G^\pm$ GCRs
were derived in Section \ref{sII}.  Noting from (\ref{so21modiden})
that the $\epsilon_\mu^+$ and $\epsilon_\mu^-$ modes are related,
we define
	\begin{equation}
	\epsilon^\mu_{n/2} = \epsilon^{\mu-}_{n/2}
	= (-1)^n \epsilon^{\mu+}_{n/2} .
	\end{equation}
Using the OPEs $\epsilon^\pm_\mu \epsilon^\mp_\nu = z^{-2/3}
g_{\mu\nu} + \ldots$ and  $\epsilon^\pm_\mu \epsilon^\pm_\nu =
z^{-1/3} {\varepsilon_{\mu\nu}}^\rho \epsilon^\mp_\rho + \ldots$,
the mode algebra for the $\epsilon^\mu$'s is found to be
\widetext
	\begin{eqnarray}
	\label{epgcr}
	\sum_{\ell=0}^\infty D^{(-\case{2}{3},-\case{1}{3})}_\ell
	\left\{ \epsilon^\mu_{\case{n-\ell}{2}}
	\epsilon^\nu_{\case{m+\ell}{2}} -
	\epsilon^\nu_{\case{m-\ell-1}{2}}
	\epsilon^\mu_{\case{n+\ell+1}{2}} \right\} &= &
	2^{-4/3} (-1)^n g^{\mu\nu} \delta_{n+m}
	\nonumber\\ &&\qquad \mbox{}		
	+ 2^{-2/3} (-1)^{n+m}
	{\varepsilon^{\mu\nu}}_\rho \epsilon^\rho_{\case{n+m}{2}} ,
	\end{eqnarray}
\narrowtext \noindent
where the coefficients $D^{(\alpha,\beta)}_\ell$ are given in
Eq.\ (\ref{Ddefn}).  Any $so(2,1)_2$ twisted-sector state can be
written as a polynomial in the $\epsilon^\mu$ creation modes acting on
the twisted-sector ground state.  The GCR (\ref{epgcr}) plus the
identity
	\begin{equation} \label{gtraceiden}
	\epsilon_{-1/2}\cdot\epsilon_0 |a\rangle = 0 ,
	\end{equation}
which is easily verified from the expression for vertex operators in
terms of the $\varphi^j$ modes (\ref{twvtx}), are sufficient to reduce
any set of such states to a linearly independent basis.

The current modes can be expressed in terms of $\epsilon^\mu$ modes as
follows.  Since the $s^+$ and $s^-$ modes are related by the
identifications (\ref{so21modiden}), we can define
	\begin{equation}
	s_{n/2} = s^-_{n/2} = (-1)^n s^+_{n/2} ,
	\end{equation}
and from the OPE $\epsilon^\pm \cdot \epsilon^\pm = 3 z^{2/3} s^\mp +
\ldots$ one then derives
\widetext
        \begin{eqnarray}
   	\label{stoep}
        s_{m + \case{1}{2}} &=& -{2^{-1/3}\over3}
	\sum_{\ell=0}^\infty D^{(-\case{5}{3},\case{2}{3})}_\ell
        \epsilon_{\case{m-\ell}{2}} \cdot
        \epsilon_{\case{m+\ell+1}{2}} , \nonumber\\
        s_m &=& {2^{-4/3}\over3}
	\sum_{\ell=0}^\infty D^{(-\case{5}{3},\case{2}{3})}_\ell
	\left\{ \epsilon_{\case{m-\ell}{2}} \cdot
	\epsilon_{\case{m+\ell}{2}} +
        \epsilon_{\case{m-\ell-1}{2}} \cdot
        \epsilon_{\case{m+\ell+1}{2}} \right\} .
        \end{eqnarray}
Similarly, since $\epsilon^\pm \cdot \epsilon^\mp = 3 z^{-2/3} +
z^{4/3} T_\varphi + \ldots$,
        \begin{equation}
	\label{ttoep}
        L^\varphi_m = {1\over8} \delta_{m,0} - (-1)^m 2^{-2/3}
	\sum_{\ell=0}^\infty D^{(\case{1}{3},-\case{7}{3})}_\ell
	\epsilon_{\case{m-\ell-1}{2}} \cdot
        \epsilon_{\case{m+\ell+1}{2}} ,
        \end{equation}
\narrowtext\noindent
where $L^\varphi_m$ are the modes of $T_\varphi$.
The mode expansion of the full stress-energy tensor is then
        \begin{equation}
   	\label{Ltwmodexp}
	L_n = {1\over 2} \sum_{\ell=-\infty}^\infty
	\alpha_{m-\ell} \cdot \alpha_\ell + L^\varphi_m .
        \end{equation}
Using (\ref{stoep}) and (\ref{ttoep}) in (\ref{Ltwmodexp}) and
(\ref{Gtwmodexp}) gives the current modes solely in terms of $\epsilon$
and $\alpha$ modes.

\subsection{Simple physical states and scattering amplitudes}

As discussed in Section \ref{sIII}B, the correct set of physical
state conditions in the $R'$ sector is not known.  So, for the
moment, we assume the maximal set:
	\begin{eqnarray}\label{pscT}
	(L_n - h_t\delta_{n,0}) |\tau\rangle &=& 0 , \nonumber\\
	(G^\pm_{n/2} - \Lambda^\pm \delta_{n,0}) |\tau\rangle &=& 0 ,
	\end{eqnarray}
for $n\ge0$ an integer.  The $h_t$ intercept determines the conformal
dimension of physical state vertex operators in the twisted sector.  In
terms of the polarization spinors of the twisted sector states, the
$L_0$ condition gives the usual Klein-Gordon equation fixing the mass
of the state.  The $G^\pm_0$ conditions, likewise, give Dirac equations
for the spinor wavefunctions, which also fix the mass of the state.
The consistency of these three zero-mode conditions determines the
$G^\pm_0$ intercepts $\Lambda^\pm$ in terms of the $L_0$ intercept
$h_t$.  We now solve these physical state conditions for the
lowest two levels of twisted-sector states.

The lowest twisted-sector state is
	\begin{equation}
	|\tau_0\rangle  = u_a|{a;k}\rangle
	\end{equation}
where $|a;k\rangle=e^{ik\cdot X}|a\rangle$ and $u_a$ is a spinor
wavefunction.  The only non-trivial physical state conditions come
from the $L_0$ and $G^\pm_0$ modes:
\widetext
	\begin{mathletters}
	\label{tauzero}
	\begin{eqnarray}
	(L_0-h_t)|\tau_0\rangle=0 \ \ &\Rightarrow&\ \
	\left(k^2 -2h_t+{1\over 4}\right) u = 0 ,\\
	(G^\pm_0-\Lambda^\pm)|\tau_0\rangle=0 \ \ &\Rightarrow&\ \
	\left(\pm i k\cdot\gamma + {3\over8} + 2^{2/3}\sqrt2
	\Lambda^\pm \right) u = 0 .
	\end{eqnarray}
	\end{mathletters}
\narrowtext \noindent
Consistency of (\ref{tauzero}a) with (\ref{tauzero}b) implies
$\Lambda^\pm$ and $h_t$ are related by
	\begin{equation}
	\label{interel}
	2^{2/3} \sqrt2 \Lambda^\pm = -{3\over8} \pm
	\sqrt{{1\over4} - 2h_t} .
	\end{equation}

Now we examine the first excited state in the twisted sector.  These
are the four dimension-5/8 states $\beta^j_{-1/2}|a\rangle$.  Written
in this manner their $so(2,1)$ properties are not apparent.  These can
be made more manifest by introducing the combinations
	\begin{equation}
	|\mu,a\rangle = \epsilon^\mu_{-1/2}|a\rangle .
	\end{equation}
{}From the identity (\ref{gtraceiden}) it follows that
	\begin{equation}
	\label{constraint}
	\gamma_\mu|\mu,a\rangle=0 .
	\end{equation}
Thus, $|\mu,a\rangle$ describes a spin-3/2 $so(2,1)$ representation.
If desired, they could be written in terms of $\varphi^j$ modes as
$|\mu,a\rangle = {\bf m}_\mu\cdot\bbox{\beta}_{-1/2} (\gamma_\mu)_b^a
|b\rangle$, (no sum on $\mu$ implied) where ${\bf m}_0 = \{-1,-1\}$,
${\bf m}_1 = \{1,0\}$, and ${\bf m}_2 = \{0,1\}$.  The general first
excited twisted-sector state is then
	\begin{equation}
	|\tau_1\rangle = u^\mu_a|{\mu,a;k}\rangle ,
	\end{equation}
where $|\mu,a;k\rangle=e^{ik\cdot X}|\mu,a\rangle$.  The constraint
(\ref{constraint}) implies that we are free to redefine the spin-3/2
polarization by $u^\mu\rightarrow u^\mu+\gamma^\mu \tilde u$, where
$\tilde u$ is an arbitrary spinor.  We fix this freedom by taking
	\begin{equation}
	\label{polconstr}
	\gamma_\mu u^\mu = 0 .
	\end{equation}

The non-trivial physical state conditions come from the $L_0$,
$G^\pm_0$ and $G^\pm_{1/2}$ modes:
\widetext
	\begin{mathletters}
	\label{tauone}
	\begin{eqnarray}
	(L_0-h_t)|\tau_1\rangle=0 \ \ &\Rightarrow&\ \
	\left( k^2 -2h_t +{5\over4}\right) u^\mu = 0 ,\\
	(G^\pm_0-\Lambda^\pm)|\tau_1\rangle=0 \ \ &\Rightarrow&
	\ \ \left(\pm ik \cdot \gamma + {5\over8}
	- 2^{2/3}\sqrt2 \Lambda^\pm \right) u^\mu = 0 ,\\
	G^\pm_{1/2}|\tau_1\rangle=0 \ \ &\Rightarrow&\ \
	k\cdot u = 0 .
	\end{eqnarray}
	\end{mathletters}
\narrowtext \noindent
Together with (\ref{polconstr}) these are the correct equations of
motion for a spin-3/2 particle.  Consistency of the Dirac equations
(\ref{tauone}b) with the $L_0$ condition (\ref{tauone}a) implies the
intercepts are related by
	\begin{equation}
	\label{interel2}
	2^{2/3}\sqrt2 \Lambda^\pm = {5\over8} \mp
	\sqrt{{5\over4} - 2h_t} .
	\end{equation}
Note that the $G^\pm_{1/2}$ condition (\ref{tauone}c) is redundant,
since it follows from the Dirac equations (\ref{tauone}b) along
with the constraint (\ref{polconstr}).

{}From these two lowest $R'$-sector levels, it already follows that the
maximal set of physical state conditions (\ref{pscT}) must be
modified.  In particular, the relations (\ref{interel}) and
(\ref{interel2}) between the $\Lambda^\pm$ and $h_t$ intercepts are
different and have no common solution.  If, however, we restrict
ourselves to the set of physical state conditions (\ref{pscRprime})
discussed in Section \ref{sIII}B and parametrized by $\alpha$, we find
that there is then a common solution for the intercepts only for
$\alpha=0$, $h_t=1/8$ and $\Lambda^+=-3\cdot2^{-25/6}$.  In other words,
a consistent set of $R'$-sector physical state conditions may be to
impose only the vanishing of the $G^+$ annihilation operators and the
$G^+_0-\Lambda^+ = 0$ condition:
	\begin{eqnarray}
	(L_n - h_t\delta_{n,0}) |\tau\rangle &=& 0 , \nonumber\\
	(G^+_{n/2} - \Lambda^+ \delta_{n,0}) |\tau\rangle &=& 0 ,
	\end{eqnarray}
for $n\ge0$ an integer.  The action of the $G^-$ operators, and
in particular the value of the $G^-_0$ intercept, would then be free to
vary from state to state as determined by the GCRs for the $c=5$ model.
With this choice of physical state conditons, the lowest-level $R'$-sector
state $\tau_0$ is a massless Majorana spinor, and the next level
describes a massive spin-3/2 particle.  Note also that since the
$G^+_{1/2}$ condition (\ref{tauone}c) is redundant, it might also be
consistent to discard the half-odd-integral modings of $G^+$ as
physical state conditions.  A potentially useful exercise would be to
compute $R'$-sector physical states at higher levels to check these
conjectures.

The simplest non-trivial scattering amplitude we can write using the
prescription (\ref{Apict2R}) developed in Section \ref{sIII} is a
three-point coupling for two $R'$-sector ground states $\tau_0$
and the massless vector state from the untwisted sector.  This
latter state was worked out for the $c=5$ model in Ref.\ \cite{T1},
and is described (in the ${\bf Z}_3$-charge $+1$ sector) by the
vertex
	\begin{equation}
	W^+(z) = \xi^\mu \left( \epsilon^+_\mu(z)
	+i{\varepsilon_\mu}^{\nu\rho} k_\nu \epsilon^+_\rho(z)
	\right) e^{ik\cdot X(z)} ,
	\end{equation}
where the momentum and polarization satisfy $k\cdot k = k\cdot\xi =
0$.  The three-point coupling is then easily worked out:
	\begin{eqnarray}
	{\cal A}_{tvt} &=& \langle \tau_0;v,k_1|
	W^+(\xi,k_2;1) |\tau_0;u,k_3 \rangle \nonumber\\
	&=& 2^{-2/3} \bar v \xi_\mu \left( \gamma^\mu
	+ i {\varepsilon^\mu}_{\nu\rho} k_2^\nu \gamma^\rho
	\right) u \,\delta^3(k_1+k_2+k_3) .
	\end{eqnarray}
The first term is just the expected minimal coupling of the fermions to
the gauge field.  The second term represents a derivative coupling
which is higher-order in the string tension, and therefore is
suppressed at energies below the Planck scale.  This string correction
to minimal coupling does not occur in the corresponding ordinary
superstring amplitude, although string correction terms do appear in
higher-point functions.

Higher-point tree amplitudes can be calculated similarly using $W^+$
vertices and $S^+=(G^+_0-\Lambda^+)^{-1}$ propagators, in accordance
with the prescription developed in Section \ref{sIII}B.  As mentioned
in that section, even without an understanding of the world-sheet
``fractional superghost'' system, the correctness of our twisted-sector
scattering prescription can still be tested at tree level by computing
four- or higher-point scattering amplitudes and checking whether
duality is satisfied.  In particular, it would be interesting to work
out some four-point amplitudes with two twisted-sector states and two
untwisted-sector states.  One could then check for duality by looking
at the spectrum of poles in the $t$-channel to see if it matched the
spectrum of the untwisted sector for some values of the twisted-sector
intercepts, while factorizing in the $s$-channel will give information
on the physical twisted-sector spectrum and may help clarify the
correct $R'$-sector physical state conditions.

\section{The $R$ sector in a \lowercase{c}=2 model}
\label{sV}

The issue of the correct set of physical state conditions in the $R$
sector is simpler than in the $R'$ sector because of the
mode identifications (\ref{ZFmodiden}) which imply that there is really
only one independent fractional supercurrent in the $R$ sector.  In
particular, since $G^-_0=\delta G^+_0$, the $G^\pm_0$ intercepts must
be related to each other by $\Lambda^- = \delta \Lambda^+$.  (Recall
that $\delta = {\rm sign}(8-c)$.) Also, by (\ref{Rsectzmalg}), the
$G^+_0$ intercept is related in a model-independent way to the $L_0$
intercept:
	\begin{equation}
	\label{RintrelGCR}
	\Lambda^+ \left( \Lambda^+ - {\delta\lambda^+\over2^{8/3}}
	\right) = {\delta\over2^{4/3}}
	\left( h_t - {5c \over 128} \right) .
	\end{equation}
This implies that
        \begin{equation}
	\label{diracompat}
	\left(G^+_0+\Lambda^+ - {\delta \lambda^+\over 2^{8/3}} \right)
        \left(G^+_0-\Lambda^+\right)
	= {\delta\over 2^{4/3}} \left(L_0- h_t\right) ,
        \end{equation}
thus ensuring that the Dirac equation resulting from the
$G^+_0-\Lambda^+=0$ physical state condition is automatically
consistent with the mass-shell constraint coming from the
$L_0-h_t=0$ physical state condition.

It is natural to ask whether the $R$-sector can be realized in the $c=5$
model described in the last section.  {}From the discussion of Section
\ref{sII}, the $R$ sector is characterized by the automorphism of the
FSC algebra which interchanges $G^+ \leftrightarrow \delta G^-$.  This
automorphism is extended to the whole $c=5$ CFT by the {\it
simultaneous} transformations $X^\mu \rightarrow -X^\mu$ and $\varphi^j
\rightarrow -\varphi^j$.  Thus the $R$ sector can be realized in the
$c=5$ model, but as the twisted-sector of a ${\bf Z}_2$ orbifold of
all five boson fields on the world-sheet.  This has unfortunate
consequences for the physical interpretation of states in this sector
since orbifolding the coordinate boson fields $X^\mu$ does not leave
the generators of space-time translations $\partial X^\mu$ invariant.
Thus there are no translation-invariant states in the $R$ sector of the
$c=5$ model.

However, one should not conclude from this that the $R$ sector is in
general badly behaved from a space-time point of view---rather, this
behavior is only a property of specific CFT models of the spin-4/3 FSC
algebra.  As an example in support of this statement, we briefly
describe in this section a $c=2$ model in which the $R$ sector appears
without orbifolding the coordinate boson fields.  Unfortunately, this
model has only one space-time dimension, so the resulting space-time
physics is trivial; however, we can still demonstrate the existence of
$R$-sector physical states which are translationally-invariant in the
one space-time dimension.

The $c=2$ model is written in terms of two free bosons, $X$ and
$\varphi$, satisfying $X(z)X(w) = -{\rm ln}(z-w)$ and $\varphi(z)
\varphi(w) = -{2\over3}{\rm ln}(z-w)$, with $\varphi$ compactified on
the unit circle $\varphi = \varphi+2\pi$ \cite{ALyT}.  The vertex
operators in the $\varphi$-CFT, $V_m = e^{im\varphi}$ for $m\in{\bf
Z}$, have conformal dimensions $h(V_m)=m^2/3$, and carry the
(untwisted-sector) ${\bf Z}_3$ charge $q=m$ mod 3.  Denoting the
dimension-1/3 and 4/3 operators by
	\begin{eqnarray}
	\epsilon^\pm &=& V_{\pm1} ,\nonumber\\
	s^\pm &=& V_{\mp2} ,
	\end{eqnarray}
it is easy to check that the spin-4/3 FSC algebra currents are given by
        \begin{equation}
	\label{c2Grep}
         G^\pm = {1\over\sqrt2} \left( i \epsilon^\pm \partial X
	+ {1\over\sqrt2} s^\pm \right) .
        \end{equation}
Comparing to the expression (\ref{Grepd}) for the currents in the $c=5$
model, the important difference for our purposes is the absence of
$\pm$ signs in front of the $\epsilon^\pm\partial X$ term in
(\ref{c2Grep}).  This means that the automorphism interchanging
$G^+\leftrightarrow G^-$ is realized in the $c=2$ CFT by the
transformation $\varphi \rightarrow -\varphi$ {\it without} any
accompanying reflection of the $X$ coordinate boson.

Thus, the $R$ sector states of the $c=2$ model are realized as states
in the twisted sector of the ${\bf Z}_2$ orbifold of the single
$\varphi$-boson.  Acting on this sector, $\varphi$ has its mode
expansion shifted in the standard way,
	\begin{equation}
	\varphi(z)=\phi+i\sum_{r\in{\bf Z}+{1\over2}}{1\over r}
	\beta_r z^{-r} ,
	\end{equation}
with modes satisfying the commutation relations $[\beta_r,\beta_s] =
(2r/3) \delta_{r+s}$.  The $\phi$ zero mode commutes with everything,
and so can be taken to be a constant, which we set to zero.  The basic
vertex operators $V_m$ have the normal ordered expansion
	\begin{eqnarray}
	V_m(z) &=& 2^{-2m^2/3} z^{-m^2/3}
	{\rm exp}\left\{-\sum_{r<0}{1\over r} m\beta_r z^{-r}\right\}
	{\rm exp}\left\{-\sum_{r>0}{1\over r} m\beta_r z^{-r}\right\} .
	\end{eqnarray}
The normal ordering factors $(4z)^{-m^2/3}$ are required to insure the
factorizability of amplitudes (associativity of the operator product).
{}From this expression acting on an arbitrary twisted-sector state
$\tau$ one obtains the bypass relation $V_m * \tau = \omega^{-m^2}
V_{-m}\,\tau$, where $\omega = e^{2\pi i/3}$.  In particular, the
single-bypass relation for the basic fields are
	\begin{eqnarray}
	\epsilon^\pm * \tau &=& \omega^2 \epsilon^\mp \tau
	\nonumber \\
	s^\pm * \tau &=& \omega^2 s^\mp \tau \nonumber \\
	G^\pm * \tau &=& \omega^2 G^\mp \tau ,
	\end{eqnarray}
the last of which is precisely the defining bypass relation
(\ref{Rbypass}) of the $R$ sector.  (More precisely, this is the bypass
relation of the $p=0$ $R$ sector; the $p=\pm1$ $R$ sectors can be
realized by letting the $\varphi$ zero mode take the values
$\phi=2\pi/3$ and $4\pi/3$.) In general, these bypass relations imply
that ${\bf Z}_2$-even fields have only integer moding, while the
${\bf Z}_2$-odd fields have only half-odd integer modings.

The twisted-sector ground state is non-degenerate.  We denote the
corresponding twist field by $\sigma(z)$ and the twist ground state by
$|\Omega\rangle = \sigma(0)|0\rangle$.  The action of the zero mode of
an arbitrary untwisted-sector operator on the twisted-sector ground
state is simply $(V_m)_0 |\Omega\rangle = 2^{-2m^2/3} |\Omega\rangle$.
Since the ground states satisfy $\beta_r|\Omega\rangle = 0$ for $r>0$,
a simple computation analogous to (\ref{dimcomp}) reveals that the
conformal dimension of the twisted-sector ground state is $h(\sigma) =
1/16$. All other twist fields are created from $\sigma$ by the repeated
action of the $\varphi$ creation modes $\beta_r$ with $0>r\in{\bf
Z}+{1\over2}$.  Thus the twist states have the spectrum of
conformal dimensions $h={1\over16}+{n\over2}$ for $n\ge0$ an integer.

Acting on a twisted-sector state, the FSC currents $G^\pm$ have the
mode expansion
	\begin{equation}
	G^\pm(z)\tau(0) = \sum_{n\in{\bf Z}} z^{-{4\over3}-{n\over2}}
	G^\pm_{n/2} z^{-n/2} \tau(0) .
	\end{equation}
Since the coordinate boson field $X$ is unaffected by the orbifolding
procedure, its mode expansion is the usual one, as in
Eq.\ (\ref{Xbosexp}).  Recalling the form of the currents
(\ref{c2Grep}), we obtain
	\begin{equation}
	G^\pm_{n/2} = {1\over\sqrt2} \left\{ \sum_{m\in{\bf Z}}
	\alpha_{m} \epsilon^\pm_{{n\over2}-m} +
	{1\over\sqrt2} s^\pm_{n\over2} \right\} .
	\end{equation}

The lowest-level state in the $R$ sector is simply
$|\tau_0\rangle = e^{ikX}|\Omega\rangle$.  The non-trivial
physical state conditions are
\widetext
	\begin{mathletters}
	\label{Rtauzero}
	\begin{eqnarray}
	(L_0-h_t)|\tau_0\rangle=0 \ \ &\Rightarrow&\ \
	k^2 -2h_t+{1\over 8} = 0 ,\\
	(G^+_0-\Lambda^+)|\tau_0\rangle=0 \ \ &\Rightarrow&\ \
	k + 2^{-5/2} - 2^{2/3}\sqrt2 \Lambda^+ = 0 .
	\end{eqnarray}
	\end{mathletters}
\narrowtext \noindent
Consistency of (\ref{Rtauzero}a) with (\ref{Rtauzero}b) implies
$\Lambda^+$ and $h_t$ are related by
	\begin{equation}
	\label{Rinterel}
	\Lambda^+ = 2^{-11/3} \pm 2^{-2/3} \sqrt{h_t-\case{1}{16}} ,
	\end{equation}
which is equivalent to the relation (\ref{RintrelGCR}) derived from the
$R$-sector mode algebra.

The first excited state in the twisted sector is $|\tau_1\rangle =
\beta_{-1/2} e^{ikX} |\Omega\rangle$.  The non-trivial physical state
conditions are:
\widetext
	\begin{mathletters}
	\label{Rtauone}
	\begin{eqnarray}
	(L_0-h_t)|\tau_1\rangle=0 \ \ &\Rightarrow&\ \
	k^2 -2h_t +{9\over8} = 0 ,\\
	(G^+_0-\Lambda^+)|\tau_1\rangle=0 \ \ &\Rightarrow& \ \
	k + 13\cdot 2^{-5/2}
	+ 3\cdot 2^{2/3}\sqrt2 \Lambda^+ = 0 ,\\
	G^+_{1/2}|\tau_1\rangle=0 \ \ &\Rightarrow&\ \
	k - 2^{-3/2} = 0 .
	\end{eqnarray}
	\end{mathletters}
\narrowtext \noindent
Consistency of (\ref{Rtauone}b) with (\ref{Rtauone}a) implies the
intercepts are related by
	\begin{equation}
	\label{Rinterel2}
	\Lambda^+ = -{1\over3}\left( 13\cdot 2^{-11/3} \pm
	2^{-2/3} \sqrt{h_t - \case{9}{16}} \right).
	\end{equation}
Although this is a different functional relation between $\Lambda^+$ and
$h_t$ than appears in (\ref{RintrelGCR}), they have the common solution
$h_t=5/8$ and $\Lambda^+=-5\cdot2^{-11/3}$, which is precisely realized
when the condition (\ref{Rtauone}c) is satisfied.

At higher levels in the $R$ sector similar physical states will be
found, all with specific values of the intercepts satisfying the
relation (\ref{RintrelGCR}).  {\it A priori}, there is no reason
to expect all these states, or even an infinite subset of them,
to have the same value of the intercept.  Of course, in one
space-time dimension duality does not require the existence
of infinite towers of states.  We have thus shown that, to the
same level of consistency, both the $R$ and $R'$ sectors can be
realized in models of the spin-4/3 fractional superstring.
It remains an open question whether either of these sectors
is actually realized in a critical ($c=10$) model of the fractional
superstring.

\acknowledgements

It is a pleasure to thank K.~Dienes, J.~Grochocinski, Z.~Kakushadze,
A.~LeClair, and E.~Witten for useful discussions and comments.  P.C.A.
would also like to thank the Center for Theoretical Physics at M.I.T.,
and B.~Zwiebach in particular, for their hospitality.  The work of
P.C.A. was supported by NSF grant PHY92-45317 and by the Ambrose Monell
Foundation, and the work of S.-H.H.T. was supported in part by the
National Science Foundation.

\appendix
\section*{Twisted-sector zero modes of $\lowercase{so}(2,1)_2$}

We take into account the action of the ${\varphi^j}$ zero modes
$\phi^j$ on the twist states.  In particular, the twist ground state
forms a representation of the nontrivial zero-mode algebra
$[\phi^1,\phi^2]=i\pi$, and so is degenerate.  We denote this
degeneracy by an index $a$ on the twist field $\sigma^a(z)$ and write
for the twist ground state simply
	\begin{equation}
	|\widetilde a\rangle = \sigma^a(0)|{0}\rangle .
	\end{equation}
We choose these states to be eigenstates of $\phi^2$ (the second
component of $\phi^j$):
	\begin{equation}\label{Bdegen}
	\phi^2 |\widetilde a\rangle = \pi a |\widetilde a\rangle ,
	\end{equation}
where, for the moment, $a$ can be any
real number.  The zero-mode algebra $[\phi^1,\phi^2] = i\pi$ implies
that $\phi^1$ and $\phi^2$ are conjugate variables, and that
	\begin{equation}
	e^{im\phi^1} |\widetilde a\rangle =
	|\widetilde{a\!+\!m}\rangle .
	\end{equation}
Recall that the classical boson
fields $\varphi^j$ take values on the torus defined by the lattice
identifications $\phi^j = \phi^j+2\pi$.  This implies, first of all,
that only exponentials ${\rm exp}(im_j\phi^j)$ with $m_j\in{\bf Z}$
should be considered (since they are single-valued on the torus), and
secondly, by (\ref{Bdegen}) that we should identify
	\begin{equation}\label{Bzmreps}
	|\widetilde{a\!+\!2}\rangle = e^{2i\phi^1} |\widetilde a\rangle
	=\beta |\widetilde a\rangle ,
	\end{equation}
where $\beta$ will be determined momentarily.  Note that since
$[e^{2i\phi^1},e^{i{\bf m}\cdot{\bbox \phi}}] = 0$ for integer $m_j$,
$\beta = e^{2i\phi^1}$ is a constant.

The fact that the $m_j$ are constrained to be integers means that for
each $0\le a<1$ and every choice of $\beta$ there is a separate,
inequivalent two-dimensional representation of the zero-mode algebra,
consisting of the states $|\widetilde a\rangle$ and
$|\widetilde{a\!+\!1}\rangle$.  Note that in terms of their $\phi^2$
eigenvalues, these two states differ by a half-lattice translation, as
do the fixed points of the ${\bf Z}_2$ orbifold.  In this way we match
up with the familiar result that the number of (ground state) twist
fields in an asymmetric (chiral) orbifold is the square-root of the
number of fixed points of the orbifold action \cite{orbs,asymm}.  Note
also that the existence of these infinite number of representations of
the zero-mode algebra is a reflection of the symmetry of the vertex
operator algebra which takes $V_{\bf m}\rightarrow
\beta^{m_1/2}\gamma^{m_2/2}V_{\bf m}$, where $\beta = e^{2i\phi^1}$ and
$\gamma = e^{2i\phi^2} = e^{2i\pi a}$.

We now show that only three of these infinite number of inequivalent
representations of the zero-mode algebra are consistent with the
$so(2,1)$ symmetry of the CFT.  In particular, the zero modes
$(U_\mu)_0$ of the generators of the $so(2,1)_2$ current algebra
symmetry must obey the $so(2,1)$ algebra
	\begin{equation}\label{Bsoalg} [(U_\mu)_0,(U_\nu)_0] =
	\varepsilon_{\mu\nu\rho}(U^\rho)_0 .  \end{equation}
Using the definitions of the $U_\mu$ fields in terms of vertices
(\ref{covflds}), and the identification of the zero mode of a vertex
acting on the ground state as
	\begin{equation}\label{Btwzmod}
	(V_{(m_1,m_2)})_0 = e^{-{i\pi\over2}m_1m_2}
	2^{-m_ig^{ij}m_j}Z_{(m_1,m_2)}, \end{equation}
from (\ref{twvtx}) where
	\begin{equation}
	Z_{(m_1,m_2)} = {\rm exp}(im_j\phi^j),
	\end{equation}
one can show,
using the Hausdorff formula $e^A e^B = e^{A+B} e^{{1\over2}[A,B]}$,
that the $so(2,1)$ algebra (\ref{Bsoalg}) is satisfied only if
	\begin{eqnarray}
	Z_{(2,1)}+Z_{(-2,-1)}+Z_{(0,3)}+Z_{(0,-3)} &=& 0
	,\nonumber\\ Z_{(1,2)}+Z_{(-1,-2)}+Z_{(3,0)}+Z_{(-3,0)} &=& 0
	,\nonumber\\ Z_{(1,-1)}+Z_{(-1,1)}+Z_{(3,3)}+Z_{(-3,-3)} &=& 0
	.  \end{eqnarray}
Acting on the ground states $|\widetilde a\rangle$, using
(\ref{Bzmreps}) and the Hausdorff formula one can show that these
equations reduce to
	\begin{eqnarray}
	\gamma+\gamma^{-2}-\beta-\beta^{-1}\gamma^{-1} &=& 0
	,\nonumber\\ \beta+\beta^{-2}-\gamma-\gamma^{-1}\beta^{-1}
	&=& 0 ,\nonumber\\
	\beta\gamma+\beta^{-2}\gamma^{-2}-\beta^{-1}-\gamma^{-1}
	&=& 0 , \end{eqnarray}
where we have defined $\gamma = e^{2i\pi a}$.  The only solutions to
these equations are $\beta = \gamma = \omega^p$ where $p\in{\bf Z}_3$
and $\omega=e^{2\pi i/3}$.  These three twisted-sector representations
give rise to three inequivalent highest-weight modules of the FSC
algebra.

Introducing the new notation $|a\rangle_p$ for the twisted sector ground
states to remove the fractional part of the index of the
$|\widetilde a\rangle$ states:
	\begin{equation}
	|a\rangle_p\equiv |\widetilde{a\!+\!\case{p}{3}}\rangle,
	\end{equation}
where $a=0$ or $1$, we can write
	\begin{eqnarray}\label{zmreps}
	e^{i{\bf m}\cdot{\bbox \phi}} |a\rangle_p & = &
	{\rm exp} \left\{{i\pi\over2}m_2\left(m_1+2a+
	{2p\over3}\right)\right\} |a+m_1\rangle_p\nonumber\\
	|a+2\rangle_p& = &\omega^p|a\rangle_p .
	\end{eqnarray}
Using the explicit form of the zero-mode representations given in
(\ref{zmreps}) and the expression (\ref{Btwzmod}) for the zero mode of
a vertex acting on the twisted-sector ground state, it is easy deduce
the zero mode actions (\ref{gammazero}) and (\ref{gammalg}).


\end{document}

\end